\documentclass[10pt,aps,prd,nofootinbib,twocolumn]{revtex4}
\usepackage[utf8]{inputenc}
\DeclareUnicodeCharacter{200B}{{\hskip 0pt}}
\usepackage{color}
\usepackage{graphicx}
\usepackage{amsmath}
\usepackage{amssymb}
\usepackage{bm}
\usepackage{acronym}
\usepackage{ifthen}
\usepackage{blindtext}
\usepackage[normalem]{ulem}
\usepackage{hyperref}
\usepackage{etoolbox}
\usepackage{fancyhdr}
\usepackage{xspace}
\usepackage{textcomp}
\usepackage{multirow}
\usepackage[caption=false]{subfig}
\usepackage{lineno}
\usepackage{tabularx}
\hypersetup{
    colorlinks=true,
    linkcolor=blue,
    filecolor=magenta,      
    urlcolor=black,
		citecolor=red,
		}
\begin{document}

\title{Memory effect for generalized modes in pp-wave spacetimes}

\author{F. L. Carneiro}\email{fernandolessa45@gmail.com}
\affiliation{Universidade Federal do Norte do Tocantins, 77824-838, Aragua\'ina, Tocantins, Brazil}

\author{H. P. de Carvalho}
\email{hyagopeixoto266@gmail.com}
\affiliation{Universidade Federal do Norte do Tocantins, 77824-838, Aragua\'ina, Tocantins, Brazil}

\author{M. P. Lobo}
\email{matheus.lobo@ufnt.edu.br
}
\affiliation{Universidade Federal do Norte do Tocantins, 77824-838, Aragua\'ina, Tocantins, Brazil}

\author{L. A. Cabral}\email{luis.cabral@ufnt.edu.br}
\affiliation{Universidade Federal do Norte do Tocantins, 77824-838, Aragua\'ina, Tocantins, Brazil}

\date{24 November 2026. Published in \href{https://doi.org/10.1103/6tnr-77jv}{Phys. Rev. D \textbf{114}, 104076 (2026)}}

\begin{abstract}
The memory effect of test particles interacting with pp-wave Gaussian pulses is investigated for { higher-order polynomial pp--wave modes beyond the standard quadrupolar $+$ and $\times$ profiles}. Massive geodesic equations are solved numerically for several values of the { polynomial index $m$}, allowing the analysis of velocity and energy memory effects. In order to eliminate possible coordinate artifacts, the study is formulated in terms of the relative motion between two test particles. The results show that the relative kinetic energy variation exhibits a quartic dependence on the wave amplitude in the regime of low initial velocities. The coefficient of this scaling is found to depend on the { transverse polynomial structure of the wave, reflecting the spatial gradients associated with the different modes}. It is further shown that the energy memory effect is determined by the integrated tidal field, linking the permanent change in the relative kinetic energy to the history of the spacetime curvature carried by the wave.
\end{abstract}


\maketitle

\section{Introduction}

General Relativity (GR) exhibits significant peculiarities when compared to other physical theories, e.g., Maxwell's electrodynamics. { In particular, its intrinsically nonlinear structure makes the analysis of radiative phenomena especially rich. Gravitational-wave memory provides an important example in which effects arising already within perturbative theory coexist with genuinely nonlinear contributions generated by the full Einstein equations.}

The memory effect is not exclusive to GR; rather, it can be observed in different physical contexts. Considering a perturbation in a material medium, one observes the presence of elastic deformations. If the perturbation carries sufficient energy, plastic deformations may occur in the material. In this case, even after the transient perturbation has passed, the configuration of the material will no longer remain the same, i.e., the material retains a memory of its interaction with the perturbation. One may visualize such memory by considering two points of the material, treated as test points, separated by a distance $l_{0}$. After the passage of the perturbation, a new distance $l$ is measured; if $l_{0} \neq l$, one may say that the material has undergone a memory effect as a result of its interaction with the perturbation.

Another widely known example occurs in the magnetization of a ferromagnetic material. In such materials, the magnetization depends not only on the instantaneous value of the applied magnetic field, but also on the entire previous history of the material's interaction with the field. This phenomenon, known as magnetic hysteresis, underlies the construction of permanent magnets.

Not only elastic and magnetic interactions leave memory in materials, but thermal interactions do so as well. For example, when iron is heated to a certain temperature and subsequently cooled, its final crystalline structure depends on the cooling history; that is, slow cooling produces a softer final state, whereas rapid cooling results in a harder state.

Even in electromagnetism, analogous effects arise in specific situations. When an electromagnetic wave interacts with a plasma, wave dissipation may occur, leading to a permanent change in the particle velocities. Thus, the plasma retains a memory of its interaction with the wave. This phenomenon is known as Landau damping \cite{dawson1961landau,thorne1991landau,manfredi1997long}.

{ Early studies of the gravitational memory effect were developed within the perturbative framework of GR. Zel'dovich and Polnarev investigated gravitational radiation emitted in encounters involving compact objects and showed that, after the passage of the radiation, freely falling test bodies may retain a permanent relative displacement \cite{zeldovich1974radiation}. Smarr subsequently investigated gravitational radiation from distant encounters and head-on collisions through the zero-frequency limit of the emitted radiation \cite{smarr1977gravitational}. These results showed that a persistent effect can already arise within perturbative gravity and is associated with a nonvanishing zero-frequency component of the radiation. This contribution is commonly referred to as ordinary or linear gravitational-wave memory, emphasizing that the existence of a memory effect does not, by itself, require nonlinear terms in the gravitational perturbation.}

{ A distinct contribution to the gravitational-wave memory was discovered by Christodoulou using the fully nonlinear Einstein equations \cite{christodoulou1991nonlinear}. In addition to the memory associated with changes in the material source, the gravitational radiation itself contributes to the permanent change of the gravitational field. Thorne subsequently provided a physical interpretation of this result, showing that the Christodoulou effect can be understood in terms of the gravitational field generated by the energy carried by the gravitational-wave burst itself \cite{thorne1992gravitational}. It is this additional contribution that is conventionally referred to as nonlinear gravitational-wave memory.}

The gravitational memory effect is traditionally characterized by a permanent relative displacement between test masses following the passage of a gravitational wave. However, this phenomenon also manifests as a permanent change in the kinematic state of the system, known as the velocity and energy memory effects. { The use of exact spacetime solutions such as plane-fronted gravitational waves (pp-waves) provides a complementary framework in which the dynamics associated with memory can be investigated beyond a perturbative description.} In this context, exploring how { different transverse structures of exact pp--waves} and integrated tidal fields dictate the permanent energy transfer to test particles is crucial for a complete understanding of the memory effect \cite{favata2010gravitational}.

These exact solutions of Einstein's equations consist of a noncaustic congruence of null geodesics, which act analogously to the light rays of plane electromagnetic waves. One of the most interesting manifestations of the nonlinearity of pp--waves is the appearance of a singularity in the collision of two sandwich waves \cite{abbasi2021probing,abbasi2023colliding,carneiro2025gravitational}.

The planar nature of the wave implies that the congruence possesses no expansion, rotation, or shear, indicating the existence of a null Killing vector $k_{\mu}$ along which the spacetime metric tensor is preserved. Using this Killing vector as the principal null vector of the congruence, i.e., its direction representing the gradient of a null coordinate $u$, and defining another null coordinate $v$, whose gradient represents the auxiliary vector $N_{\mu}$, one may describe the line element of such a spacetime, in doubly null coordinates $x^{\mu}=(u,x,y,v)$, as
\begin{equation}\label{eq1}
ds^{2}=H(u,x,y)\,du^{2}+dx^{2}+dy^{2}-2\,du\,dv\,,
\end{equation}
where the coordinates $x$ and $y$ describe the space transverse to the direction of wave propagation. These are known as Brinkmann coordinates \cite{brinkmann1925einstein}.
The null coordinates are related to Cartesian ones by
\begin{equation}\label{coordtransformation}
	u=\frac{t-z}{\sqrt{2}}\,,\qquad v=\frac{t+z}{\sqrt{2}}\,.
\end{equation}

The function $H(u,x,y)$ must satisfy the two-dimensional Laplace equation
\begin{equation}
(\partial_{x}^{2}+\partial_{y}^{2})H(u,x,y)=0\,,
\end{equation}
whose general solution may be written as the real (or imaginary) part of a Laurent series \cite{podolsky2014gyratonic},
\begin{equation}\label{eq3}
F(u,\zeta) =
\sum_{m=1}^{\infty}
\left( a_{m}(u)\,\zeta^{m} + b_{m}(u)\,\zeta^{-m} \right)
+\mu(u)\ln\zeta\,,
\end{equation}
with $\zeta=x+iy$ representing a complex coordinate defined on the plane perpendicular to the direction of wave propagation. Einstein's equations do not determine the $u$-dependence of the function, reflecting the freedom in the choice of the wave pulse profile. { Different choices of the profile function lead to distinct geometric classes of pp--wave spacetimes.} Several classification schemes based on conformal and homothetic vector fields have been proposed, both in GR and in modified gravity theories, revealing a structure of symmetry classes for these spacetimes \cite{hussain2019classification,hussain2021conformal}.

Plane gravitational waves possess a rich symmetry structure that goes beyond ordinary isometries. In particular, they admit conformal and chronoprojective symmetries that generate additional conserved quantities for particle dynamics in these backgrounds \cite{zhang2020scaling}. Such symmetries are closely related to the correspondence between gravitational waves and time-dependent oscillators, which can be established through Niederer-type transformations \cite{andrzejewski2019niederer}.

{ Accordingly, the solutions \eqref{eq3} determine different transverse structures of the pp--wave profile, obtained by taking either the real or imaginary part of the function with respect to $\zeta$ and assuming $u$ real, i.e., $H(u,x,y)=\mathrm{Re}[F(u,\zeta)]$ or $H(u,x,y)=\mathrm{Im}[F(u,\zeta)]$.} The logarithmic term and the coefficients $b_{m}$ do not represent vacuum solutions, since the Einstein tensor becomes singular along the propagation axis of the wave, corresponding to monopole and multipole solutions propagating together with the wave.

For $m=1$, the term proportional to $a_{m}$ represents a function that can be removed by an appropriate coordinate transformation, generating a continuous and uniform displacement in a system of particles, without geodesic deviation. For $m=2$, one obtains
\begin{align}
H_{+}(u,x,y)&=f(u)(x^{2}-y^{2})\,,\label{plus}\\
H_{\times}(u,x,y)&=f(u)(x\,y)\,.\label{times}
\end{align}

These polarization modes are so named because of the distortion profile they produce in a circular ring of particles. These polarization states correspond to the two polarization modes found in the solutions of gravitational waves derived from the linearized Einstein equations \cite{d2022introducing}. 
Interestingly, similar dynamical structures appear in other physical systems. For instance, it has been shown that the dynamics of particles in periodic plane gravitational waves can be mapped to ion trap systems, providing a useful analogy between gravitational radiation and anisotropic oscillator models \cite{zhang2018ion}.

We observe geodesic separation between test particles during the passage of the wave. However, even when considering a pulse of finite duration, e.g., a Gaussian pulse, the separation between the particles does not return to its initial value after the wave has passed; therefore, the particle system retains a memory of its interaction with the wave, i.e., a memory effect. Mathematically, this phenomenon may be interpreted as a nontrivial holonomy associated with a closed path in spacetime, with the separation vectors failing to return to their initial configuration after the dynamical cycle. This geometric interpretation provides a unified perspective on the memory effect, connecting it to the global structure of spacetime and to the asymptotic conditions imposed by gravitational radiation.

Investigations of the memory effect in pp-waves are carried out by numerically solving the massive geodesic equations for two test particles $A$ and $B$, computing their positions $\vec{r}_{A}$ and $\vec{r}_{B}$ before and after the passage of the wave, and comparing the initial separation $\Delta r_{i}$ with the final separation $\Delta r_{f}$. Analogously to plastic deformation in materials, if $\Delta r_{i}\neq\Delta r_{f}$, a memory of the interaction with the gravitational wave persists. In recent years, this research program has been developed in the literature \cite{zhang2017memory,zhang2018memory,datta2024memory}, culminating in the identification of the so-called velocity memory effect, in which not only the relative position vector between the particles is permanently altered, but also the relative velocity vector \cite{zhang2018velocity,zhang2024displacement,zhao2025gravitational}. A natural extension of these results leads to the concept of an energy memory effect, since a permanent change in particle velocity implies a lasting variation in their energy \cite{maluf2018kinetic,maluf2018plane,maluf2018variations}. Similar energy exchanges between gravitational waves and free particles have also been investigated in other wave geometries, where variations in kinetic energy and angular momentum arise from the interaction with the gravitational field \cite{abbasi2022study,abbasi2025kinetic}.

A complementary perspective arises in the study of impulsive gravitational waves obtained through ultrarelativistic boosting procedures, where the spacetime curvature becomes distributional and is supported on null hypersurfaces. In such cases, the interaction with test particles leads to discontinuities and refraction effects in their trajectories across the wavefront, reflecting an instantaneous transfer of gravitational influence \cite{battista2016riemann}.

{ Despite this progressive development and the growing interest in the memory effect, most analyses of particle memory in pp--wave backgrounds have focused on the standard quadratic profiles associated with the $+$ and $\times$ polarizations. The general solution \eqref{eq3}, however, also admits regular polynomial terms with $m>2$, which define exact pp--wave geometries with increasingly nonuniform transverse structure. Unlike the $m=2$ case, for which the curvature is independent of the transverse coordinates and the spacetime belongs to the plane-wave subclass, the curvature of the $m>2$ profiles depends explicitly on the transverse position.}

{ It is important to emphasize that the index $m$ appearing in the polynomial profiles $\zeta^{m}$ characterizes the angular and radial dependence of the wave on the two-dimensional transverse plane. It should not be directly identified with the $(\ell,m)$ indices of a spherical-harmonic decomposition of gravitational radiation generated by an asymptotically flat astrophysical source. In particular, for the regular polynomial modes considered here, $H\sim r^{m}$ at large transverse distance, so that the $m>2$ profiles are not intended as global far-zone models of realistic astrophysical waveforms. Rather, in the present work they are treated as exact pp--wave geometries whose different transverse tidal structures provide a theoretical setting in which the response and memory of massive test particles can be investigated. Polynomial pp--wave profiles of this type have also recently attracted attention in other theoretical contexts, such as chaotic geodesic motion and escape-basin structures \cite{rossetto2026wada}.}

{ The absence of a presently established source mechanism producing these particular higher-order polynomial profiles does not, by itself, exclude a possible phenomenological relevance. Establishing such a connection would require an independent analysis relating these exact geometries to source-generated gravitational-wave configurations, which is beyond the scope of the present work. Accordingly, our purpose here is not to identify the polynomial index $m$ with astrophysical multipolar radiation, but to characterize how the transverse structure of these exact pp--waves is imprinted on the memory of test particles.}

{ In this paper, we investigate this question by analyzing the memory effect of particles interacting with several polynomial pp--wave profiles.} By considering Gaussian profiles, we evaluate the velocity memory effect through the permanent change in kinetic energy in the asymptotic regime and find that, for low speeds $v\sim 10^{-4}$, the particles' relative energy variation as a function of the wave amplitude follows a quartic behavior, whose coefficient is sensitive to the mode $m$. We also show that particles can gain or lose kinetic energy, depending on their initial conditions and on the amplitude of the wave (for all values of $m$ considered here), with a tendency to gain energy when their velocity is small and to lose it when their velocity is large, resembling the Landau damping effect. Moreover, when interacting with GWs of large amplitude, particles tend to gain energy, whereas they tend to lose it when the amplitude is small. Hence, we speculate that the qualitative behavior (energy gain or loss) depends on the relative configuration between the wave and the particle, rather than on the wave or the particle individually. 

In addition, our results indicate that the energy memory effect admits a direct interpretation in terms of the spacetime curvature. In particular, the variation of the relative kinetic energy can be related to the square of a memory tensor defined as the integral of the tidal components of the Riemann tensor along the particle worldlines. This establishes that the energy transfer is governed by the integrated curvature of the wave, revealing its nonlocal character. { The higher-order dependence of the relative kinetic-energy variation on the wave amplitude discussed in this work should therefore be understood as a property of the particle dynamics in the exact pp-wave geometry. In particular, it should not be identified with the nonlinear gravitational-wave memory in the sense of Christodoulou, which has a distinct physical origin as described above.}

Our article is divided as follows. In Sec.~\ref{sec2}, we explore the \eqref{plus} and \eqref{times} modes, including their relation to linearized gravitational waves and their memory effect. In Sec.~\ref{sec3}, we pursue our main goals by considering the kinetic energy variation of the particles for several values of $m$, analyzing the dependence of the energy gain on the wave amplitude and the effect of the inhomogeneity of the Riemann tensor (present for $m>2$) in the transverse plane on the relative acceleration of two particles, i.e., on the tidal forces of spacetime. In Sec.~\ref{conc}, we present our conclusions. We use the geometrized unit system $c=G=1$ and the $+2$ signature.

\section{Memory effect for $+$ and $\times$ modes}\label{sec2}

In this section, we analyze the memory effect associated with the standard polarization modes of plane gravitational waves. We begin by examining the relation between the exact pp-wave solutions and the weak-field limit of gravitational waves, showing how the familiar $+$ and $\times$ polarizations arise in the linearized approximation. { This discussion also illustrates explicitly how a permanent displacement may already arise within the linearized description of the gravitational field.}

\subsection{Linearized limit}

By choosing the polarization \eqref{plus} in the line element \eqref{eq1} and defining the new coordinates
{
\begin{align}
x &= g(u)X\,, \nonumber\\
y &= h(u)Y\,, \nonumber\\
v &= V+\frac{1}{2}\left(\dot{g}gX^{2}+\dot{h}hY^{2}\right)\,,
\label{transformation}
\end{align}
the line element can be written in the Rosen form
\begin{equation}\label{rosenplus}
	ds^{2}=-2dudV+g(u)^{2}dX^{2}+h(u)^{2}dY^{2}\,,
\end{equation}
provided that
\begin{equation}
\ddot{g}=fg\,,\qquad \ddot{h}=-fh\,.
\end{equation}
These conditions arise from requiring the cancellation of the remaining $du^{2}$ terms after the coordinate transformation.}
The dot indicates differentiation with respect to $u$. The line element \eqref{rosenplus} represents the pp-wave spacetime in the so-called Rosen coordinates.
{ To make contact with the weak-field description, we consider small deviations of the transverse metric functions from their Minkowski values,
\begin{equation*}
g(u)=1+\delta g(u),\qquad
h(u)=1+\delta h(u),
\qquad
|\delta g|,|\delta h|\ll1 .
\end{equation*}
Retaining only terms linear in these deviations, the metric becomes
\begin{equation}\label{linear}
ds^2=-2dudV+[1+2\delta g(u)]dX^2+[1+2\delta h(u)]dY^2,
\end{equation}
which corresponds to the weak-field form of a gravitational wave in Rosen coordinates.
}

A similar construction may be performed for the $\times$ polarization. In this case, the wave profile couples the two transverse coordinates, and therefore the transformation to Rosen coordinates must involve linear combinations of the new transverse coordinates rather than simple rescalings. After performing an appropriate coordinate transformation and redefining the null coordinate $v$ in order to eliminate mixed terms involving $du$, the metric can again be written in Rosen form, where the transverse metric depends only on the null coordinate $u$. In contrast with the $+$ polarization, the resulting transverse metric generally contains a mixed term involving both transverse directions, reflecting the characteristic distortion pattern of the $\times$ polarization. In the weak-field limit, this metric reduces to the usual linearized gravitational wave with $\times$ polarization, showing that both descriptions correspond to the same physical degrees of freedom, differing only by a rotation in the transverse plane.

When we consider $m=1$, the ring undergoes only a translation without deformation, i.e., there is no geodesic deviation. This feature corresponds merely to a coordinate transformation that may appear as a physical motion, whereas the genuine gravitational effect arises from geodesic deviation.

{ We may see the emergence of the memory effect from \eqref{linear}. Consider two freely falling particles with constant Rosen-coordinate separation $\Delta X$ along the $X$ direction. To first order in the metric perturbation, their physical separation is
\begin{equation}
L_X(u)=g(u)\Delta X
      \simeq [1+\delta g(u)]\Delta X.
\end{equation}
The net change between the asymptotic regions is therefore
\begin{equation}\label{linearmemory}
\Delta L_X
=
L_X(+\infty)-L_X(-\infty)
\simeq
\left[\delta g(+\infty)-\delta g(-\infty)\right]\Delta X.
\end{equation}
Consequently, whenever the metric perturbation approaches different values before and after the passage of the wave, a permanent relative displacement remains already at linear order.}

{ As a simple example, one may choose
\begin{equation}
\delta g(u)=-\delta h(u)
=\epsilon\,[\tanh(u)-1]\,,
\qquad |\epsilon|\ll1.
\end{equation}
In this case,
\begin{equation}
\delta g(-\infty)=-2\epsilon,
\qquad
\delta g(+\infty)=0,
\end{equation}
and Eq.~\eqref{linearmemory} gives
\begin{equation}
\Delta L_X\simeq2\epsilon\,\Delta X\neq0.
\end{equation}
Hence, we may see the occurrence of a memory effect, since the particles' distance is permanently altered.}

\subsection{Polarization and tidal interpretation}

We proceed to investigate the memory effect in pp-waves through their interaction with a massive timelike test particle. In principle, one may employ any coordinate system. While Rosen coordinates provide some connection with the linearized theory, Brinkmann coordinates are richer in describing more generalized modes. Hence, we use the line element \eqref{eq1} throughout this article.

In order to develop an intuition regarding the polarization modes, we consider the relative acceleration between two nearby geodesics governed by the geodesic deviation equation
\begin{equation}
\frac{D^{2}\xi^{\mu}}{d\tau^{2}}=-R^{\mu}{}_{\alpha\nu\beta}\,u^{\alpha}u^{\beta}\,\xi^{\nu},
\end{equation}
where $u^{\mu}$ is the four-velocity of the reference geodesic and $\xi^{\mu}$ is the separation vector between neighboring particles. For gravitational waves propagating along the null coordinate $u$, the nonzero components of the Riemann tensor are
\begin{align}
	R^{i}{}_{uju}&=-\frac{1}{2}\partial_{i}\partial_{j}H\,,\nonumber\\
		R^{v}{}_{iju}&=-\frac{1}{2}\partial_{i}\partial_{j}H\label{riemanngeneral}\,,
\end{align}
where $i,j=1,2$. 

The tidal forces responsible for the deformation of a ring of test particles are fully encoded in the second derivatives of $H$. For the quadratic profiles corresponding to the $m=2$ modes, these derivatives are constant in the transverse coordinates, producing a uniform tidal field across the plane. For higher modes, however, the derivatives acquire spatial dependence, leading to nonuniform tidal forces and more intricate patterns of geodesic deviation.

For the $+$ polarization, we have
\begin{equation}
R^{x}{}_{uxu}=-R^{y}{}_{uyu}=-f(u),
\qquad
R^{x}{}_{uyu}=0 .
\end{equation}
In this case, the tidal field stretches the separation between particles along one transverse direction while compressing it along the orthogonal direction. As the wave propagates, the ring of test particles is alternately elongated and contracted along the $x$ and $y$ axes, producing a deformation pattern that resembles the symbol ``$+$'', which motivates the name of this polarization.

For the $\times$ polarization, the relevant components become
\begin{equation}
R^{x}{}_{uyu}=R^{y}{}_{uxu}=-\frac{1}{2}f(u),
\qquad
R^{x}{}_{uxu}=R^{y}{}_{uyu}=0 .
\end{equation}
In this case, the tidal forces act along directions rotated by $45^\circ$ with respect to the coordinate axes. Consequently, a circular ring of particles is deformed along two diagonal directions of the transverse plane, producing a pattern that resembles the symbol ``$\times$''. This geometrical deformation provides the physical origin of the naming of the two polarization states of gravitational waves.

It is worth emphasizing that in Brinkmann coordinates the tidal field is  directly determined by the second derivatives of the profile function $H$, as can be seen in Eq.~\eqref{riemanngeneral}. Consequently, the physical content of the gravitational wave is entirely encoded in the curvature tensor, independently of coordinate choices in the transverse plane.

\subsection{Numerical setup}

{
The memory effect can be perceived when considering a finite pulse, e.g., a Gaussian profile for the function $f(u)$. If we consider the particle ring before and after the interaction (in the asymptotic limit), its pattern is not the same, i.e., a perfect circle does not remain a circle after interacting with the wave, thus retaining a memory of the interaction.

Let $\tau$ denote the proper time along a timelike geodesic. The geodesic equation for the null coordinate $u$ gives
\begin{equation}
\frac{d^{2}u}{d\tau^{2}}=0,
\end{equation}
and therefore
\begin{equation}\label{kappa}
\frac{du}{d\tau}=\kappa,
\end{equation}
where $\kappa$ is a constant along the trajectory. We keep $\kappa$ arbitrary at this stage, since its value is related to the physical initial state of the particle. Since $u=\kappa\tau+{\rm const.}$, for $\kappa\neq0$ the coordinate $u$ itself may be used as an affine parameter for the numerical integration. In what follows, the dot denotes differentiation with respect to $u$.

In terms of this parameter, the geodesic equations are
\begin{align}
\ddot{x} - \frac{1}{2}\,\partial_{x}H&=0\,,\label{geoX}\\
\ddot{y} - \frac{1}{2}\,\partial_{y}H&=0\,,\label{geoY}\\
\ddot{v} -  \frac{1}{2}\partial_{u}H
        - (\partial_{x}H)\dot{x}
        - (\partial_{y}H)\dot{y} &=0\,.\label{geoV}
\end{align}
Thus, the transverse equations used in the numerical integration are independent of the particular value of $\kappa$. While we may work directly with the null coordinate $v$, we find it more convenient to use the Cartesian-like coordinate $z$. Hence, using \eqref{coordtransformation}, Eq.~\eqref{geoV} becomes
\begin{equation}\label{geoZ}
\ddot{z}-\frac{1}{2\sqrt{2}}\,\partial_{u}H
-\frac{1}{\sqrt{2}}(\partial_{x}H)\,\dot{x}
-\frac{1}{\sqrt{2}}(\partial_{y}H)\,\dot{y}=0.
\end{equation}

For an arbitrary profile $f(u)$, the system formed by the geodesic Eqs.~\eqref{geoX}--\eqref{geoV} cannot, in general, be solved analytically.
Therefore, we resort to numerical integration. In principle, six initial
conditions must be specified for the spatial trajectory. However, for a
fixed value of $\kappa$, the corresponding velocity components are not
completely independent.

The four--velocity, defined with respect to the proper time,
\begin{equation}
U^{\mu}=\frac{dx^{\mu}}{d\tau},
\end{equation}
satisfies the normalization condition
\begin{equation}
g_{\mu\nu}U^{\mu}U^{\nu}=-1.
\end{equation}
Using Eq.~\eqref{kappa}, one has
\begin{equation}
U^{\mu}
=
\kappa\,\frac{dx^{\mu}}{du}.
\end{equation}
Consequently, the tangent vector associated with the $u$ parametrization satisfies
\begin{equation}\label{normu}
g_{\mu\nu}
\frac{dx^{\mu}}{du}
\frac{dx^{\nu}}{du}
=
-\frac{1}{\kappa^{2}}.
\end{equation}
For the Brinkmann line element \eqref{eq1}, Eq.~\eqref{normu} yields
\begin{equation}
H+\dot{x}^{2}+\dot{y}^{2}-2\dot{v}
=
-\frac{1}{\kappa^{2}},
\end{equation}
and therefore
\begin{equation}\label{dotv}
\dot{v}
=
\frac{1}{2}
\left(
H+\dot{x}^{2}+\dot{y}^{2}
+\frac{1}{\kappa^{2}}
\right).
\end{equation}
Thus, once $\kappa$ and the transverse motion are specified, the null component $\dot v$ is fixed by the normalization condition.

Again using \eqref{coordtransformation}, we obtain
\begin{equation}\label{geoZnormal}
\dot{z}
=
\frac{\dot{v}-1}{\sqrt{2}}
=
\frac{1}{2\sqrt{2}}
\left(
H+\dot{x}^{2}+\dot{y}^{2}
+\frac{1}{\kappa^{2}}-2
\right),
\end{equation}
showing that, for a fixed $\kappa$, the longitudinal motion is not an independent degree of freedom, but is determined by the transverse motion and the spacetime geometry.

The physical three--velocity must be defined with respect to the Cartesian-like time coordinate $t$. From \eqref{coordtransformation},
\begin{equation}
\dot{t}
=
\frac{1+\dot{v}}{\sqrt{2}},
\qquad
\dot{z}
=
\frac{\dot{v}-1}{\sqrt{2}},
\end{equation}
and hence
\begin{align}
\frac{dx}{dt}
&=
\frac{\sqrt{2}\,\dot{x}}{1+\dot{v}},\label{physicalvelocityX}
\\
\frac{dy}{dt}
&=
\frac{\sqrt{2}\,\dot{y}}{1+\dot{v}},
\\
\frac{dz}{dt}
&=
\frac{\dot{v}-1}{\dot{v}+1}.
\label{physicalvelocity}
\end{align}

The value of $\kappa$ must therefore be chosen consistently with the physical initial state under consideration. In particular, consider a particle initially at rest in an asymptotically Minkowskian region, where $H=0$. The conditions
\begin{equation}
\frac{dx}{dt}
=
\frac{dy}{dt}
=
\frac{dz}{dt}
=0
\end{equation}
imply
\begin{equation}
\dot{x}=\dot{y}=0,
\qquad
\dot{v}=1.
\end{equation}
Eq.~\eqref{dotv} then gives
\begin{equation}
1=\frac{1}{2\kappa^{2}},
\end{equation}
so that
\begin{equation}\label{kapparest}
\kappa=\frac{1}{\sqrt{2}}.
\end{equation}
Thus, Eq.~\eqref{kapparest} corresponds to a particle initially at rest when the transverse velocities vanish in the asymptotically Minkowskian region. Unless otherwise stated, this parametrization is adopted in all numerical examples and figures throughout the remainder of the article.

More generally, in the asymptotic Minkowski region the physical meaning of $\kappa$ can be made explicit by writing
\begin{equation}
\kappa
=
\frac{du}{d\tau}
=
\frac{1}{\sqrt{2}}
\left(
\frac{dt}{d\tau}
-
\frac{dz}{d\tau}
\right)
=
\frac{\gamma}{\sqrt{2}}
\left(1-v_{z}\right),
\end{equation}
where
\begin{equation}
\gamma=
\frac{1}{\sqrt{1-v_{x}^{2}-v_{y}^{2}-v_{z}^{2}}}.
\end{equation}
Hence, different choices of $\kappa$ correspond to different physical initial states of the particle.
}
\subsection{Velocity and energy memory}

We begin by illustrating the $+$ and $\times$ patterns of geodesic deviation mentioned above. For this purpose, we consider a Gaussian wave profile
\begin{equation}\label{gaussian}
	f(u)=A\, e^{-u^{2}/\Lambda^{2}},
\end{equation}
where the constants $A$ and $\Lambda$ fix the dimensions of the function and are related to the amplitude and the characteristic width of the pulse, respectively. 

\begin{figure}[t]
\centering
\begin{minipage}{0.48\linewidth}
\centering
\includegraphics[width=\linewidth]{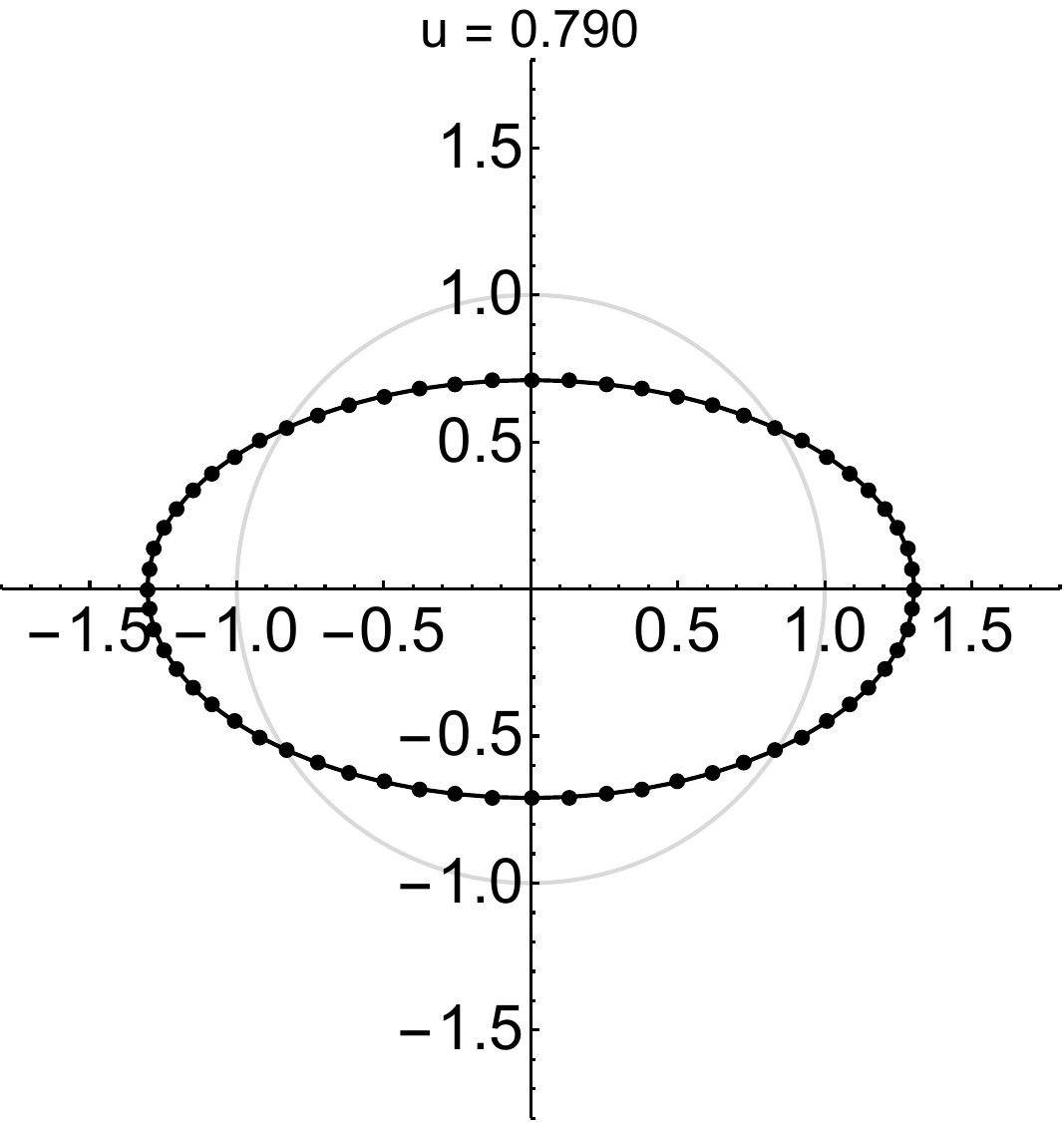}
\par\smallskip
{\small (a) Deformation produced by a pp-wave with $+$ polarization.}
\end{minipage}\hfill
\begin{minipage}{0.48\linewidth}
\centering
\includegraphics[width=\linewidth]{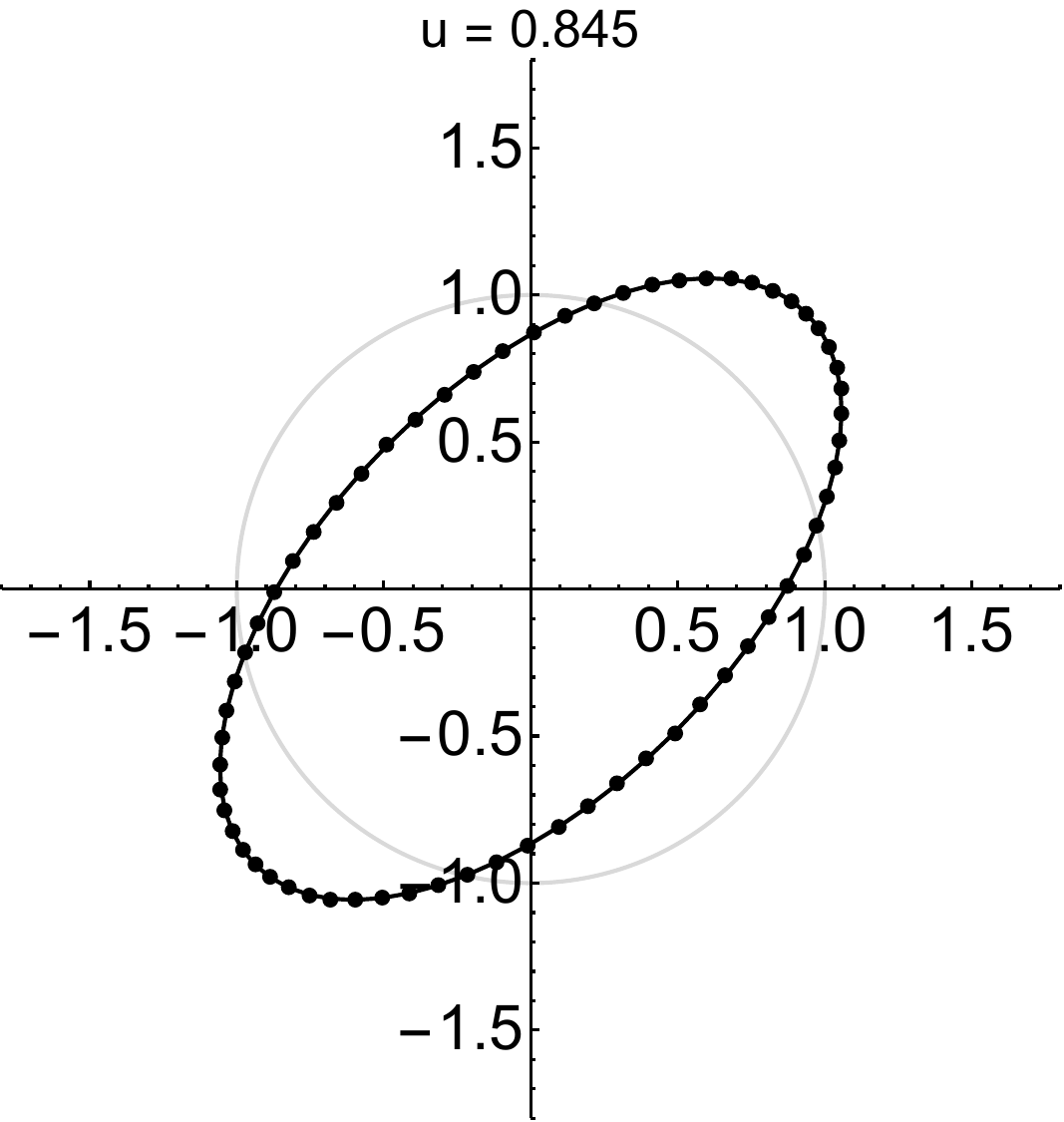}
\par\smallskip
{\small (b) Deformation produced by a pp-wave with $\times$ polarization.}
\end{minipage}
\caption{Deformation of a particle ring of radius $R=1$ caused by a pp-wave of mode $m=2$. The wave parameters are $A=1$ and $\Lambda=0.15\sqrt{2}$.}
\label{fig1}
\end{figure}

We consider a set of particles initially at rest arranged along a perfect circle of radius $R=1$. By numerically integrating Eqs.~\eqref{geoX} and \eqref{geoY}, the deformation patterns produced by the two polarization states can be obtained. The resulting configurations after an elapsed value of the coordinate $u$ are shown in Fig.~\ref{fig1}.  
The ring consists of $64$ particles uniformly distributed along the circle. All particles are initially at rest, i.e., $\dot{x}_{0}=0=\dot{y}_{0}$, { which also implies $\dot z_{0}=0$,} where the subscript $0$ denotes the instant before the passage of the wave. In principle this corresponds to $u=-\infty$, but for numerical purposes we take $u=-10$ as the initial time.

\begin{figure}[t]
\centering
\includegraphics[width=\linewidth]{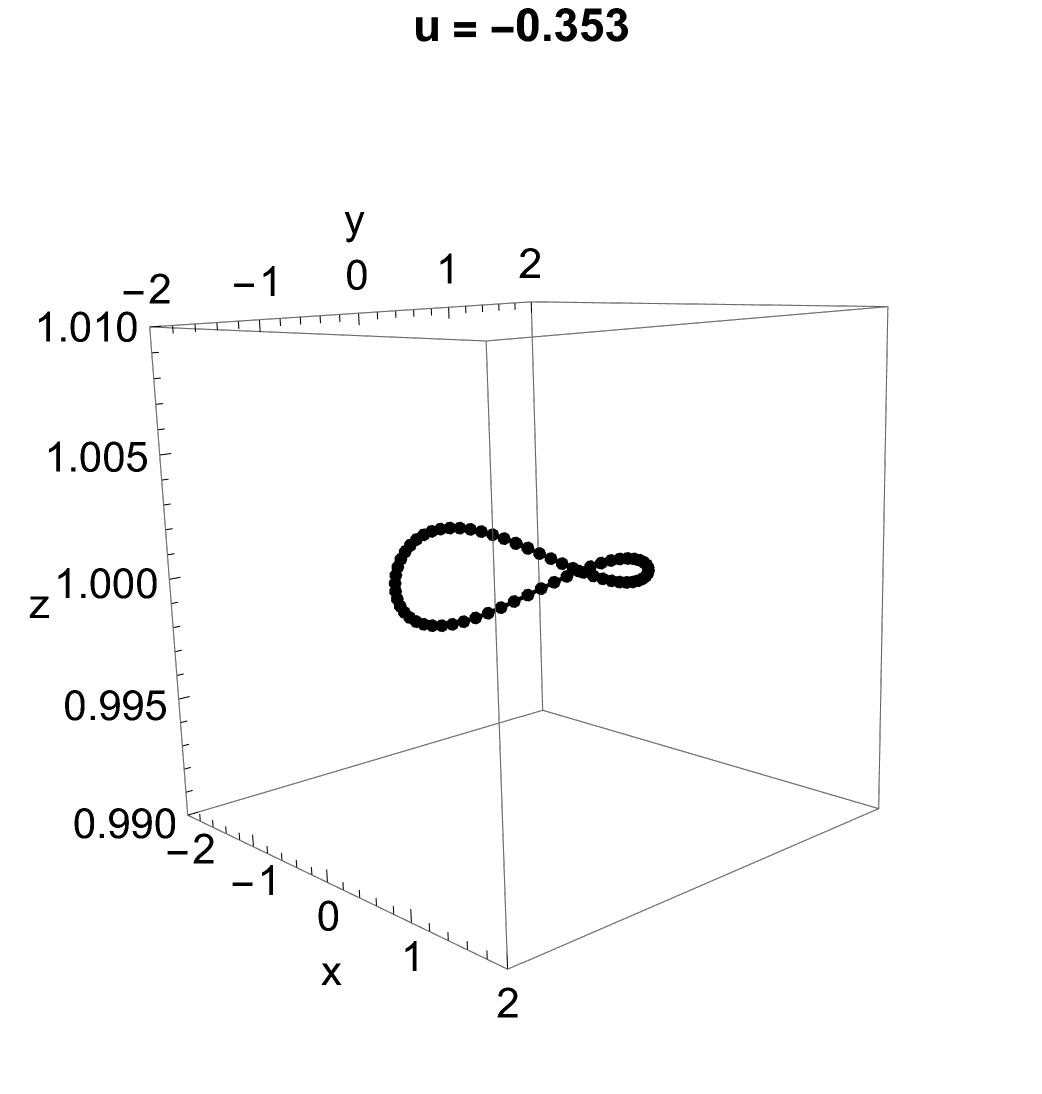}
\caption{Out-of-plane deformation of a particle ring of radius $R=1$ caused by a pp-wave with $+$ polarization. The wave parameters are $A=1$ and $\Lambda=0.15\sqrt{2}$.}
\label{fig2}
\end{figure}

While we have displayed only a transverse view, the gravitational wave also distorts the circle longitudinally, bending it, as can be seen in Fig.~\ref{fig2}.

Even after the wave influence ceases, the particles keep moving with constant speed, never returning to rest. This feature is known as the velocity memory effect. The presence of the velocity memory effect raises an interesting question: if a free particle is initially at rest, its kinetic energy is zero, but the permanent change in its velocity causes the particle to acquire kinetic energy, retaining it after the interaction. Thus, we may say that a pp-wave also exhibits an energy memory effect. This effect may be interpreted as a permanent transfer of energy between the gravitational wave and the particle motion.

We may compute the energy memory effect by considering two conditions: (i) low velocities; (ii) the asymptotic limit. In such regimes, the spacetime becomes approximately Minkowskian and the low-velocity approximation applies, so we expect Newtonian physics to hold. { Since the dot denotes differentiation with respect to $u$, the kinetic energy must be expressed in terms of the physical coordinate velocities with respect to $t$. Hence, using Eqs.~\eqref{physicalvelocityX}--\eqref{physicalvelocity}, we define the kinetic energy per unit mass as
\begin{align}
K&=\frac{1}{2}\left[
\left(\frac{dx}{dt}\right)^{2}
+\left(\frac{dy}{dt}\right)^{2}
+\left(\frac{dz}{dt}\right)^{2}
\right]\nonumber\\
&=\frac{1}{2}\left[
\frac{2(\dot{x}^{2}+\dot{y}^{2})}{(1+\dot v)^{2}}
+\left(\frac{\dot v-1}{\dot v+1}\right)^{2}
\right].
\end{align}
Defining
\begin{equation}
S=\dot{x}^{2}+\dot{y}^{2},
\end{equation}
and using Eq.~\eqref{dotv}, this expression becomes
\begin{equation}\label{kinetic}
K=
\frac{
8S+
\left(
H+S+\kappa^{-2}-2
\right)^{2}}
{2\left(
H+S+\kappa^{-2}+2
\right)^{2}}.
\end{equation}}

{ The interpretation of Eq.~\eqref{kinetic} as the usual kinetic energy applies to the asymptotically Minkowskian regions before and after the passage of the wave, where $H\rightarrow0$. For the numerical examples considered in this section, using the parametrization \eqref{kapparest}, Eq.~\eqref{kinetic} reduces asymptotically to
\begin{equation}\label{kineticapprox}
K\approx
\frac{
S(S+8)}
{2(S+4)^{2}}.
\end{equation}
In the low-velocity regime, $S\ll1$, one further obtains
\begin{equation}
K\approx\frac{S}{4}.
\end{equation}
The factor $1/4$ reflects the relation between derivatives with respect to $u$ and the physical velocities with respect to $t$ under the parametrization \eqref{kapparest}.}

{ For a fixed value of $\kappa$,} we see that the kinetic energy, and consequently its variation, depends only on the transverse motion { in the asymptotic regions}, since Eqs.~\eqref{geoX} and \eqref{geoY} depend only on the transverse coordinates through $H(u,x,y)$. Therefore, { once $\kappa$ is fixed,} only the initial conditions $x_{0},y_{0},\dot{x}_{0},\dot{y}_{0}$ are relevant for the velocity memory effect, reducing the number of controlled variables, which was not considered in Refs.~\cite{maluf2018kinetic,maluf2018plane,maluf2018variations}.

\begin{figure}[t]
\centering
\begin{minipage}{0.48\linewidth}
\centering
\includegraphics[width=\linewidth]{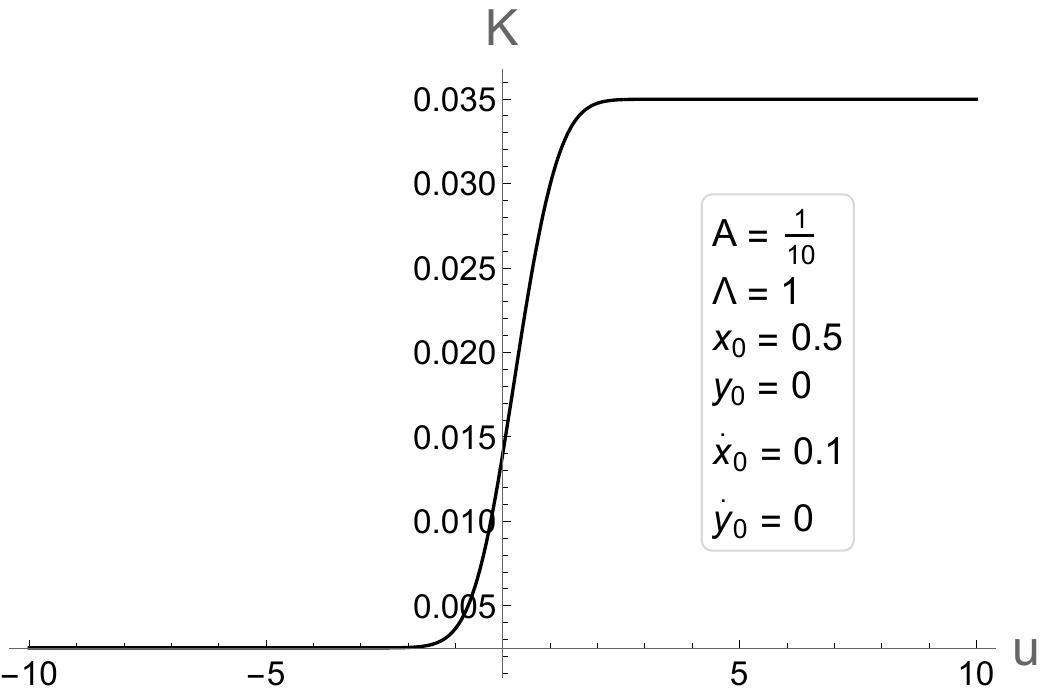}
\par\smallskip
{\small (a) Kinetic energy gain.}
\end{minipage}\hfill
\begin{minipage}{0.48\linewidth}
\centering
\includegraphics[width=\linewidth]{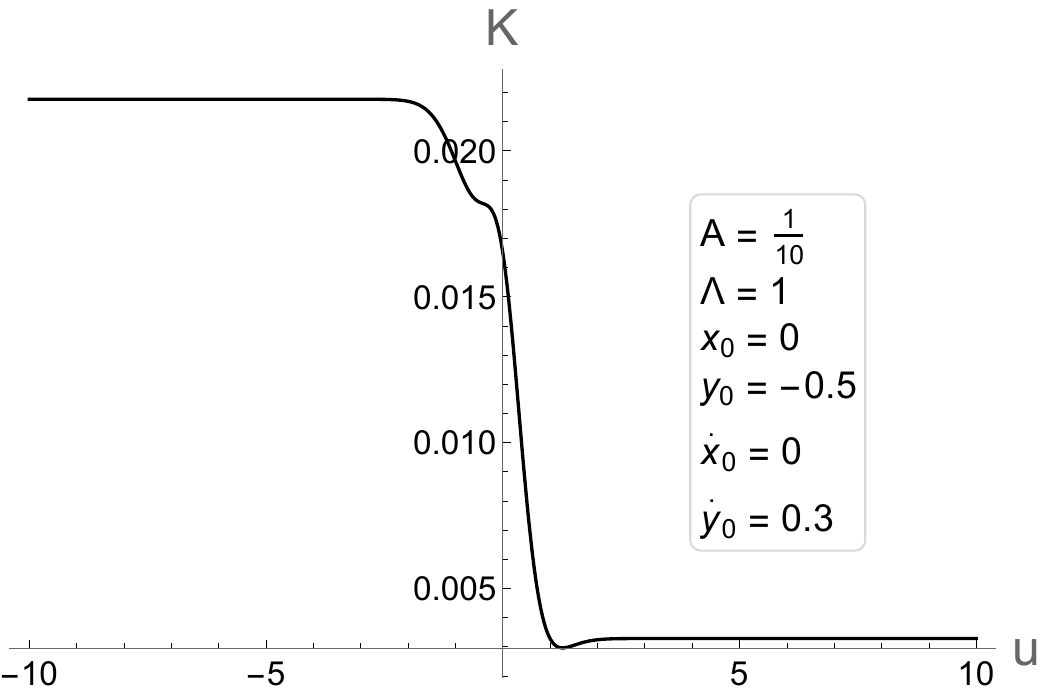}
\par\smallskip
{\small (b) Kinetic energy loss.}
\end{minipage}
\caption{{ Kinetic energy \eqref{kinetic} variation for a pp-wave with $+$ polarization.}}
\label{fig3}
\end{figure}

If the particle is initially at rest, it is clear that it can only gain energy. But what happens if it already possesses nonzero kinetic energy? { We now consider particles with nonvanishing transverse initial velocities.} Let us consider some examples shown in Fig.~\ref{fig3}. For the initial conditions presented in Fig.~\ref{fig3}(a), the particle gains kinetic energy when interacting with the wave. However, for the same wave, another particle with a different set of initial conditions loses energy, as shown in Fig.~\ref{fig3}(b). Two conclusions can be drawn: (i) a particle can lose energy to the gravitational wave; (ii) the qualitative behavior (gain or loss) depends on the particle initial conditions.

The same qualitative behavior is observed for a pp-wave with $\times$ polarization.

\begin{figure}[t]
\centering
\includegraphics[width=\linewidth]{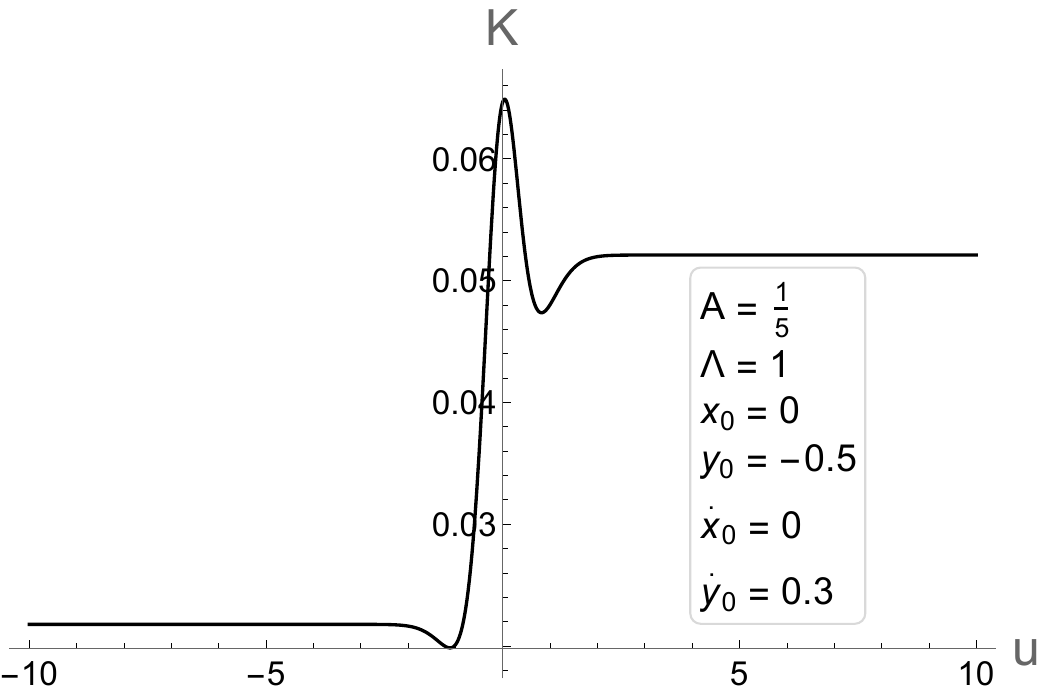}
\caption{{ Kinetic energy variation for the same particle displayed in Fig.~\ref{fig3}(b), but for another amplitude.}}
\label{fig4}
\end{figure}

Let us now consider the same particle of Fig.~\ref{fig3}(b) interacting with a different wave, as illustrated in Fig.~\ref{fig4}. In this case, the same initial conditions that previously led to an energy loss now lead to an energy gain. Hence, we may infer a third conclusion: (iii) the same particle may gain or lose energy when interacting with different gravitational waves.

By combining conclusions (ii) and (iii), we conclude that the energy gain/loss behavior depends on the relative configuration between the particle and the gravitational wave. A similar feature has been observed in analyses based on adapted Fermi coordinates, where memory effects also depend on combined parameters of the wave and the particle, rather than on either individually \cite{achour2024displacement}.

One might argue that these results lack physical significance, since they could be a spurious effect of coordinate transformations or be restricted to the particular polarizations \eqref{plus} and \eqref{times}. In the next section we address these issues by considering the relative motion between two particles and { additional polynomial modes}.

\section{Memory effect for higher-order modes}\label{sec3}

{ Higher-order polynomial modes} introduce a richer transverse structure in pp--waves, leading to increasingly intricate tidal patterns in the motion of test particles. In this section we investigate how these higher modes affect the gravitational memory effect, both in the deformation of particle configurations and in the transfer of kinetic energy between the wave and the particles.

\subsection{Geodesic patterns for higher modes}

\begin{figure}[t]
\centering
\begin{minipage}{0.48\linewidth}
\centering
\includegraphics[width=\linewidth]{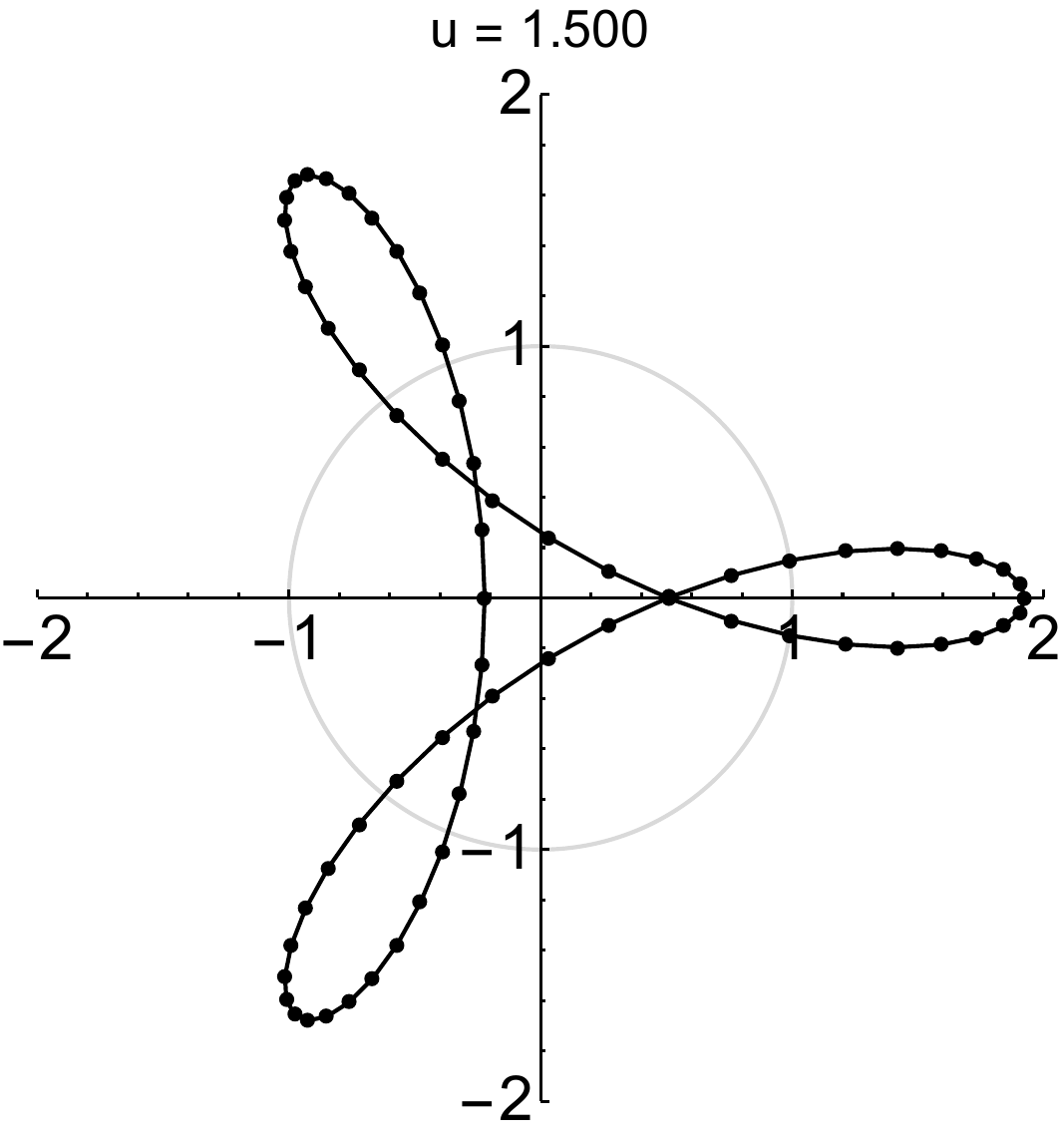}
\par\smallskip
{\small (a) Deformation produced by a pp-wave with the { $H_{3R}$ profile} \eqref{H3R}.}
\end{minipage}\hfill
\begin{minipage}{0.48\linewidth}
\centering
\includegraphics[width=\linewidth]{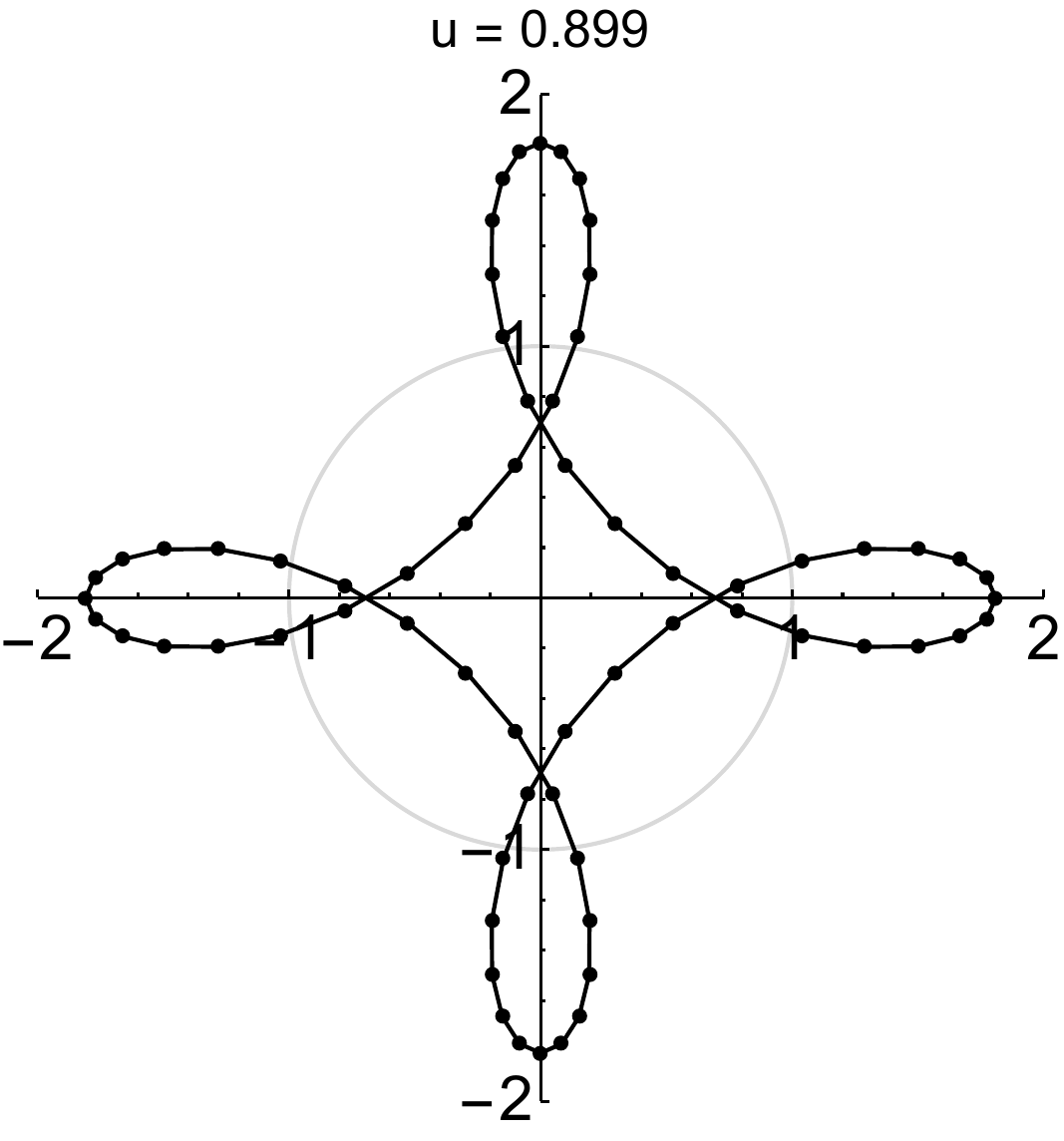}
\par\smallskip
{\small (b) Deformation produced by a pp-wave with the { $H_{4R}$ profile} \eqref{H4R}.}
\end{minipage}
\caption{Deformation of a particle ring of radius $R=1$ caused by pp-waves of higher modes. The wave parameters are $A=1$ and $\Lambda=0.15\sqrt{2}$.}
\label{fig5}
\end{figure}

When considering the geodesic motion for higher modes one might initially expect the motion to appear chaotic. However, interesting patterns emerge when we consider the same initial ring of particles initially at rest used in Fig.~\ref{fig1}. As can be seen in Figs.~\ref{fig1} and \ref{fig5}, for each mode $m$ there are $m$ points where the ring initially stretches. For $m=2$ we had two points, producing an ellipse. For 
\begin{align}
	H_{3R}(u,x,y)&=f(u)\,Re(\zeta^{3})=f(u)(x^{3}-3xy^{2})\,,\label{H3R}\\
	H_{4R}(u,x,y)&=f(u)\,Re(\zeta^{4})=f(u)(x^{4}-6x^{2}y^{2}+y^{4})\,,\label{H4R}
\end{align}
we obtain floral-like patterns where the number of petals corresponds to the value of $m$. This geometrical interpretation reflects the { polynomial transverse structure} encoded in the transverse derivatives of the profile function $H$. Similar patterns appear for higher values of $m$. When considering the { corresponding imaginary polynomial modes}, i.e., by taking the imaginary parts, we obtain the same behavior but rotated by $\pi/2m$ in the transverse plane.

This pattern may be understood from the tidal matrix $\mathcal{T}_{ij}=-\frac{1}{2}\partial_i\partial_j H$, whose eigenvectors determine the principal directions of stretching and compression on the transverse plane. For $m=2$ these directions are uniform, leading to the familiar quadrupolar deformation. For higher modes, however, the Hessian of $H$ varies with the angular position, so that the principal tidal directions rotate along the ring. The resulting alternation of stretching sectors gives rise to the floral patterns with $m$ petals.

We see that the behavior becomes more intricate than in the case $m=2$, where stretching occurs along one direction and compression along the orthogonal one. In the higher modes, different portions of the ring alternately approach and separate from each other while moving relative to other regions that evolve in a similar manner.

\begin{figure}[t]
\centering
\begin{minipage}{0.48\linewidth}
\centering
\includegraphics[width=\linewidth]{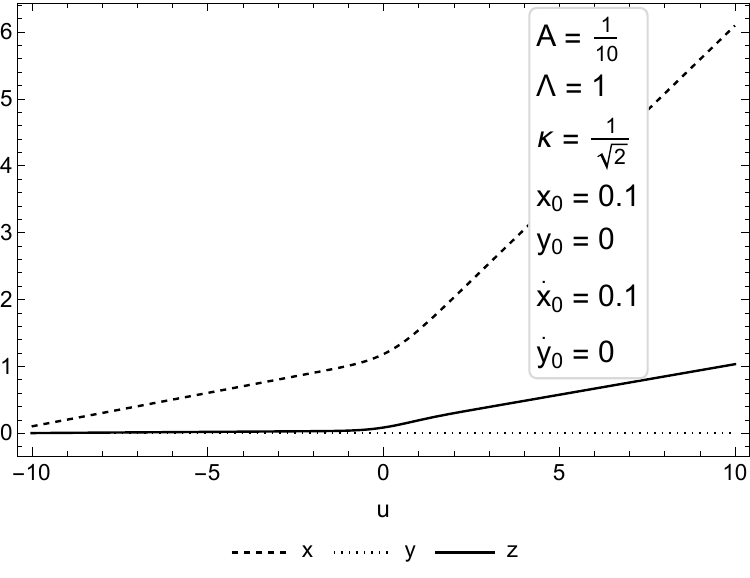}
\par\smallskip
{\small (a) Coordinates.}
\end{minipage}\hfill
\begin{minipage}{0.48\linewidth}
\centering
\includegraphics[width=\linewidth]{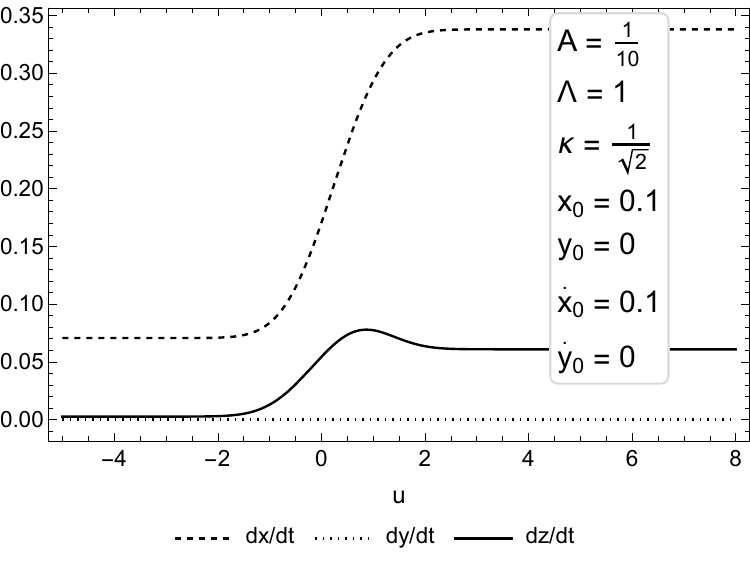}
\par\smallskip
{\small { (b) Physical coordinate velocities.}}
\end{minipage}
\caption{Evolution of the motion for the interaction with a pp--wave with the { $H_{3R}$ profile} \eqref{H3R}. { Panel (b) displays $dx/dt$, $dy/dt$, and $dz/dt$.}}
\label{fig6}
\end{figure}

\begin{figure}[t]
\centering
\begin{minipage}{0.48\linewidth}
\centering
\includegraphics[width=\linewidth]{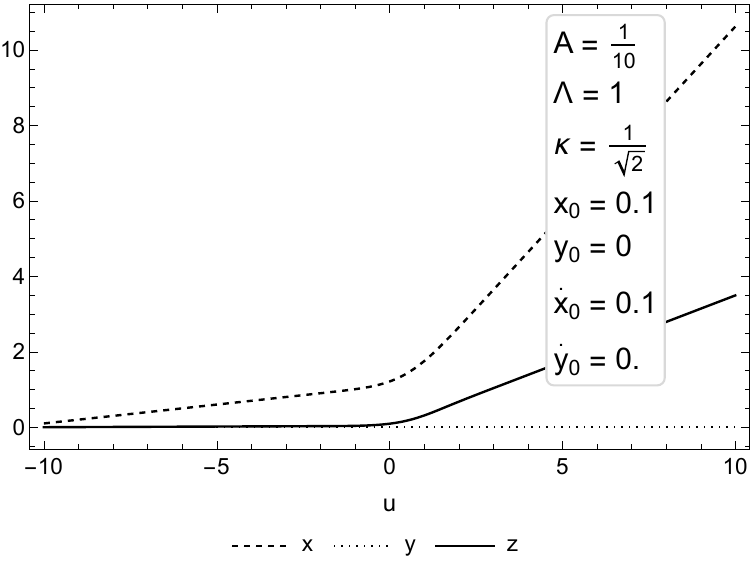}
\par\smallskip
{\small (a) Coordinates.}
\end{minipage}\hfill
\begin{minipage}{0.48\linewidth}
\centering
\includegraphics[width=\linewidth]{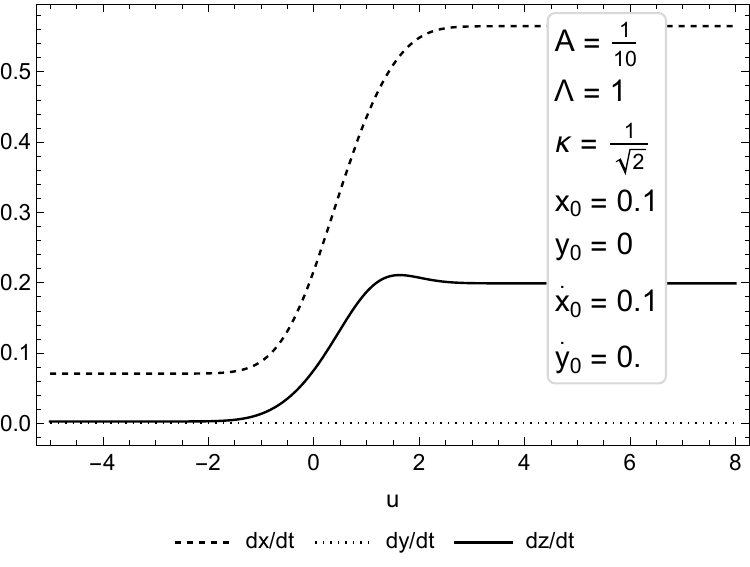}
\par\smallskip
{\small { (b) Physical coordinate velocities.}}
\end{minipage}
\caption{Evolution of the motion for the interaction with a pp--wave with the { $H_{4R}$ profile} \eqref{H4R}. { Panel (b) displays $dx/dt$, $dy/dt$, and $dz/dt$.}}
\label{fig7}
\end{figure}

One may then ask whether, in this more complex scenario, the behavior of the memory effect changes. By analyzing the geodesic motion in the same way as in the previous section, we observe that the memory effect persists. { The transverse coordinates are obtained from Eqs.~\eqref{geoX} and \eqref{geoY}, whereas the longitudinal coordinate is determined consistently with the normalization condition through Eq.~\eqref{geoZnormal}. Since the dot denotes differentiation with respect to $u$, the velocity panels display the physical coordinate velocities $dx/dt$, $dy/dt$, and $dz/dt$, as defined in Eqs.~\eqref{physicalvelocityX}--\eqref{physicalvelocity}, rather than the corresponding derivatives with respect to $u$.} This can be seen in Fig.~\ref{fig6}, which displays the coordinates and { physical coordinate velocities} of a test particle for the { $H_{3R}$ profile}, and in Fig.~\ref{fig7} for the case $H_{4R}$. The same behavior is observed for higher modes. Regarding the energy memory effect, we again observe a dependence on the initial conditions and on the wave amplitude for the qualitative behavior, although with distinct quantitative differences.

\subsection{Relative kinetic energy and tidal interpretation}

{
In order to mitigate the possibility that the memory effect in the kinetic energy is merely associated with the behavior of a single trajectory in a particular coordinate system, we proceed differently from Refs.~\cite{maluf2018plane,maluf2018kinetic,maluf2018variations} by considering the relative motion between two particles. Let us therefore consider two particles, labelled by $A$ and $B$, initially placed on a ring of radius $R$ and separated by an angular distance $\Delta\theta$.

Since the dot denotes differentiation with respect to the affine parameter $u$, the physical coordinate velocities must be evaluated with respect to the Cartesian-like time coordinate $t$, as discussed in Sec.~II C. For each particle $P=A,B$, let us define
\begin{equation}
S_P=\dot{x}_P^{\,2}+\dot{y}_P^{\,2},
\end{equation}
together with
\begin{align}
D_P&=H_P+S_P+\kappa^{-2}+2,\\
N_P&=H_P+S_P+\kappa^{-2}-2,
\end{align}
where $H_P=H(u,x_P,y_P)$. Using Eqs.~\eqref{physicalvelocityX}--\eqref{physicalvelocity}, the physical coordinate velocities of each particle are
\begin{align}
V_P^x&\equiv\frac{dx_P}{dt}
=\frac{2\sqrt{2}\,\dot{x}_P}{D_P},\\
V_P^y&\equiv\frac{dy_P}{dt}
=\frac{2\sqrt{2}\,\dot{y}_P}{D_P},\\
V_P^z&\equiv\frac{dz_P}{dt}
=\frac{N_P}{D_P}.
\end{align}

The relative squared velocity is then defined, in this common Cartesian-like frame, as
\begin{equation}
v_{\rm rel}^{2}
=
(V_B^x-V_A^x)^2
+
(V_B^y-V_A^y)^2
+
(V_B^z-V_A^z)^2,
\label{vrel}
\end{equation}
and the corresponding relative kinetic energy per unit mass is
\begin{equation}
K_{\rm rel}
=
\frac{1}{2}v_{\rm rel}^{2}.
\label{Kreldef}
\end{equation}
This quantity will be evaluated in the asymptotically Minkowskian regions, where its interpretation as a classical relative kinetic energy is appropriate in the low-velocity regime.

Before and after the passage of the Gaussian pulse, one has $H_P\rightarrow0$. Therefore,
\begin{align}
D_P^{\rm asy}&=S_P+\kappa^{-2}+2,\\
N_P^{\rm asy}&=S_P+\kappa^{-2}-2,
\end{align}
and Eq.~\eqref{Kreldef} becomes
\begin{align}
K_{\rm rel}^{\rm asy}
=
\frac{1}{2}
\Bigg\{
&8\left[
\frac{\dot{x}_B}{D_B^{\rm asy}}
-
\frac{\dot{x}_A}{D_A^{\rm asy}}
\right]^2
+
8\left[
\frac{\dot{y}_B}{D_B^{\rm asy}}
-
\frac{\dot{y}_A}{D_A^{\rm asy}}
\right]^2
\nonumber\\
&+
\left[
\frac{N_B^{\rm asy}}{D_B^{\rm asy}}
-
\frac{N_A^{\rm asy}}{D_A^{\rm asy}}
\right]^2
\Bigg\}.
\label{Krelasym}
\end{align}

For the numerical configurations considered below, we specialize to the parametrization \eqref{kapparest}. In this case,
\begin{equation}
D_P^{\rm asy}=S_P+4,
\qquad
N_P^{\rm asy}=S_P,
\end{equation}
so that
\begin{align}
K_{\rm rel}^{\rm asy}
=
\frac{1}{2}
\Bigg\{
&8\left[
\frac{\dot{x}_B}{S_B+4}
-
\frac{\dot{x}_A}{S_A+4}
\right]^2
+
8\left[
\frac{\dot{y}_B}{S_B+4}
-
\frac{\dot{y}_A}{S_A+4}
\right]^2
\nonumber\\
&+
\left[
\frac{S_B}{S_B+4}
-
\frac{S_A}{S_A+4}
\right]^2
\Bigg\}.
\label{Krelkappa}
\end{align}

The corresponding memory effect in the relative kinetic energy is defined by
\begin{equation}\label{relativeenergy}
\Delta K_{\rm rel}
=
K_{\rm rel}^{f}-K_{\rm rel}^{i},
\end{equation}
where the superscripts $i$ and $f$ denote the asymptotic regions before and after the interaction with the wave.

The advantage of considering $K_{\rm rel}$ instead of the kinetic energy of a single particle is that the former directly probes the permanent modification of the relative motion between the two trajectories. In the asymptotically Minkowskian regions, this allows the comparison to be performed in a common inertial frame. For neighboring particles, the relation between this quantity and the tidal action of the gravitational field can be made explicit through the geodesic deviation equation.

Let us denote by
\begin{equation}
\xi^{i}=x_{B}^{i}-x_{A}^{i},
\qquad i=1,2,
\end{equation}
the transverse separation vector between two neighboring particles. Since the dot denotes differentiation with respect to $u$,
\begin{equation}
\dot{\xi}^{i}
=
\dot{x}_{B}^{i}-\dot{x}_{A}^{i}
\end{equation}
represents the relative transverse derivative with respect to the affine parameter $u$, rather than directly the physical relative velocity.

For pp--waves, the nontrivial tidal action takes place in the transverse plane, and the geodesic deviation equation expressed in terms of the affine parameter $u$ reads
\begin{equation}
\frac{d^{2}\xi^{i}}{du^{2}}
=
-\,R^{i}{}_{uju}\,\xi^{j},
\qquad i,j=1,2.
\label{geodevtrans}
\end{equation}
From Eq.~\eqref{riemanngeneral}, one sees explicitly that the relative transverse motion is governed by the gravitational tidal field.

Integrating Eq.~\eqref{geodevtrans} from the remote past to the remote future gives
\begin{equation}
\Delta\dot{\xi}^{i}
=
\dot{\xi}^{i}(+\infty)-\dot{\xi}^{i}(-\infty)
=
-\int_{-\infty}^{+\infty}
R^{i}{}_{uju}(u)\,\xi^{j}(u)\,du.
\label{exactmemory}
\end{equation}
Eq.~\eqref{exactmemory} is exact within the geodesic-deviation description. It shows that the final relative transverse motion depends on the full history of the tidal field during the passage of the wave, namely on the curvature integrated along the trajectory and weighted by the instantaneous separation vector $\xi^{j}(u)$. Thus, the memory is not determined only by the local value of the curvature at one instant, but by the cumulative tidal action over the whole pulse.

An important subtlety is that the separation vector $\xi^{j}(u)$ itself evolves during the interaction and therefore cannot, in general, be taken outside the integral in Eq.~\eqref{exactmemory}. However, if the pulse is sufficiently short, or if the separation varies only mildly while the wave is passing, one may approximate
\begin{equation}
\xi^{j}(u)\approx\xi^{j}_{0},
\end{equation}
where $\xi^{j}_{0}$ denotes the initial separation vector. Under this assumption,
\begin{equation}
\Delta\dot{\xi}^{i}
\approx
-\left(
\int_{-\infty}^{+\infty}
R^{i}{}_{uju}(u)\,du
\right)\xi^{j}_{0}.
\end{equation}

This motivates the definition of the integrated tidal tensor, or memory tensor,
\begin{equation}
\mathcal{M}^{i}{}_{j}
=
-\int_{-\infty}^{+\infty}
R^{i}{}_{uju}(u)\,du,
\label{memorytensor}
\end{equation}
such that
\begin{equation}
\Delta\dot{\xi}^{i}
\approx
\mathcal{M}^{i}{}_{j}\xi^{j}_{0}.
\label{linearizedmemory}
\end{equation}
The tensor $\mathcal{M}^{i}{}_{j}$ therefore characterizes the integrated tidal action of the wave on the transverse plane and maps, within the approximation above, the initial separation vector into the permanent change of its derivative with respect to $u$.

For the regular polynomial profiles considered here, the transverse dependence may be written as
\begin{equation}\label{Hform}
H(u,x,y)=f(u)\,\Phi_m(x,y),
\end{equation}
where, for instance,
\begin{equation}
\Phi_m(x,y)=\mathrm{Re}[(x+iy)^m]\,.
\end{equation}
Since
\begin{equation}
R^{i}{}_{uju}
=
-\frac{1}{2}\partial^{i}\partial_{j}H,
\end{equation}
the transverse structure of the memory tensor is governed by the Hessian of $\Phi_m$. Different polynomial modes therefore generate different integrated tidal maps in the transverse plane, which are reflected in the relative motion of the particles.

To connect Eq.~\eqref{linearizedmemory} with the physical relative velocity, we now consider the asymptotic low-velocity regime and the parametrization \eqref{kapparest} used in our numerical analysis. In this regime,
\begin{equation}
\frac{dx^{i}}{dt}
\simeq
\kappa\,\dot{x}^{i},
\end{equation}
and therefore the transverse relative physical velocity satisfies
\begin{equation}
V_{\rm rel}^{i}
\simeq
\kappa\,\dot{\xi}^{i}.
\end{equation}
Consequently,
\begin{equation}
\Delta V_{\rm rel}^{i}
\simeq
\kappa\,
\mathcal{M}^{i}{}_{j}\xi^{j}_{0}.
\label{physicalmemory}
\end{equation}

The transverse contribution to the relative kinetic energy in this regime is
\begin{equation}
K_{\rm rel}^{\perp}
=
\frac{1}{2}\delta_{ij}
V_{\rm rel}^{i}V_{\rm rel}^{j}.
\end{equation}
In the general case, where the particles may already possess an initial relative velocity, its variation is
\begin{align}
\Delta K_{\rm rel}^{\perp}
\simeq\,
&
\kappa\,
\delta_{ij}
V_{{\rm rel},i}^{\,i}
\left(
\mathcal{M}^{j}{}_{k}\xi^{k}_{0}
\right)
\nonumber\\
&+
\frac{\kappa^{2}}{2}
\delta_{ij}
\left(
\mathcal{M}^{i}{}_{k}\xi^{k}_{0}
\right)
\left(
\mathcal{M}^{j}{}_{\ell}\xi^{\ell}_{0}
\right),
\label{DKgeneral}
\end{align}
where $V_{{\rm rel},i}^{\,i}$ denotes the initial transverse relative physical velocity.

For particles that are initially comoving,
\begin{equation}
V_{{\rm rel},i}^{\,i}=0,
\end{equation}
and Eq.~\eqref{DKgeneral} reduces to
\begin{equation}
\Delta K_{\rm rel}^{\perp}
\simeq
\frac{\kappa^{2}}{2}
\xi^{k}_{0}\xi^{\ell}_{0}
\mathcal{M}^{i}{}_{k}\mathcal{M}_{i\ell}.
\label{DKM}
\end{equation}
Using the parametrization \eqref{kapparest},
\begin{equation}
\Delta K_{\rm rel}^{\perp}
\simeq
\frac{1}{4}
\xi^{k}_{0}\xi^{\ell}_{0}
\mathcal{M}^{i}{}_{k}\mathcal{M}_{i\ell}.
\end{equation}

Thus, for initially comoving particles, the transverse relative kinetic-energy memory is quadratic in the integrated tidal tensor. More generally, Eq.~\eqref{DKgeneral} shows that the energy variation is controlled by the integrated curvature together with the initial relative state of the particles. In either case, the permanent energy change depends on the history of the gravitational tidal field carried by the wave.

In this sense, the use of $\Delta K_{\rm rel}$ provides a clearer physical characterization than the kinetic energy of an isolated trajectory. Rather than monitoring only the behavior of a single worldline, it probes the permanent modification of the relative motion of freely falling particles in the asymptotically Minkowskian regions. Its connection with the geodesic deviation equation makes explicit that this persistent change is governed by the integrated tidal action of the pp--wave.
}

\subsection{Numerical dependence on the wave amplitude}

{
We aim to investigate quantitatively how the different modes affect the energy memory effect. Hence, we compute Eq.~\eqref{relativeenergy} for hundreds of different values of the wave amplitude $A$ and for several modes $m$, obtaining a relation
\begin{equation}
\Delta K_{\rm rel}=\Delta K_{\rm rel}(A).
\end{equation}
Throughout this subsection, we evaluate the relative kinetic energy from the physical coordinate velocities according to Eqs.~\eqref{Kreldef} and \eqref{Krelkappa}.

For $m=2$, the results are shown in Fig.~\ref{fig8}. We choose $\theta_{A}=\pi$, $\theta_{B}=0$, and $R=1$, so that the two particles are initially located at the leftmost and rightmost points of the ring, respectively. We also choose $\Lambda=1$.

\begin{figure}[t]
\centering
\includegraphics[width=\linewidth]{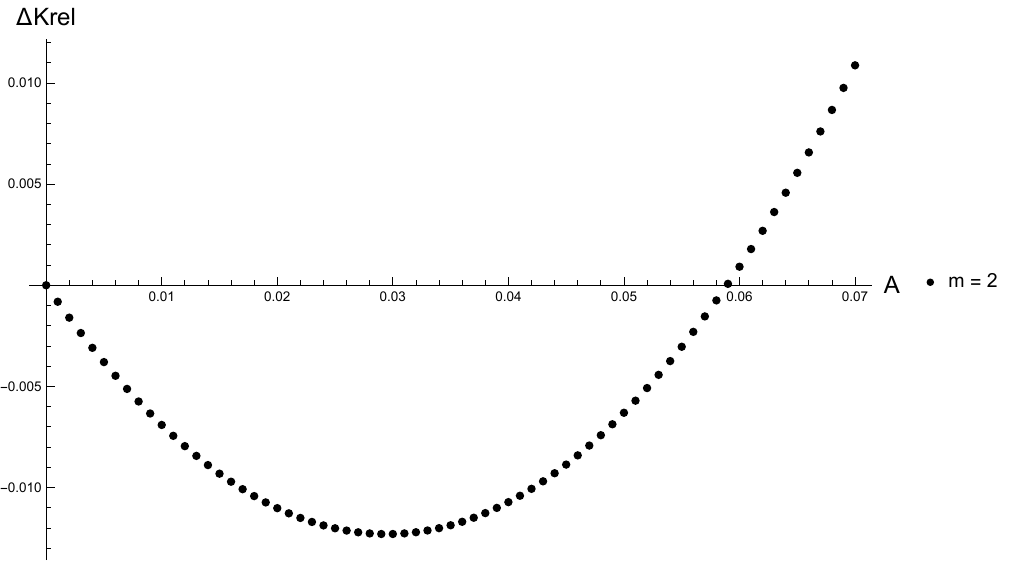}
\caption{{ Relative kinetic energy variation for the $+$ polarization. The initial transverse derivatives with respect to $u$ are $\dot{x}_{A0}=\dot{y}_{A0}=-0.1$, $\dot{x}_{B0}=0$, and $\dot{y}_{B0}=0.3$.}}
\label{fig8}
\end{figure}

From Fig.~\ref{fig8} we clearly observe that the sign of $\Delta K_{\rm rel}$ depends on the amplitude of the wave. Thus, even for fixed initial positions and transverse derivatives, the same pair of particles may exhibit either an increase or a decrease in relative kinetic energy depending on the wave amplitude. In Ref.~\cite{zhang2025displacement}, the authors also found a notable relation between the amplitude and the energy, although considering distinct pulse profiles and a particle initially at rest.

Proceeding to the analysis of higher modes, we now consider $\theta_{A}=0$ and $\theta_{B}=0.01$, corresponding to two neighboring particles on the initial ring. The results are shown in Fig.~\ref{fig9}, where the numerical points are connected in order to improve the visualization.

\begin{figure}[t]
\centering
\includegraphics[width=\linewidth]{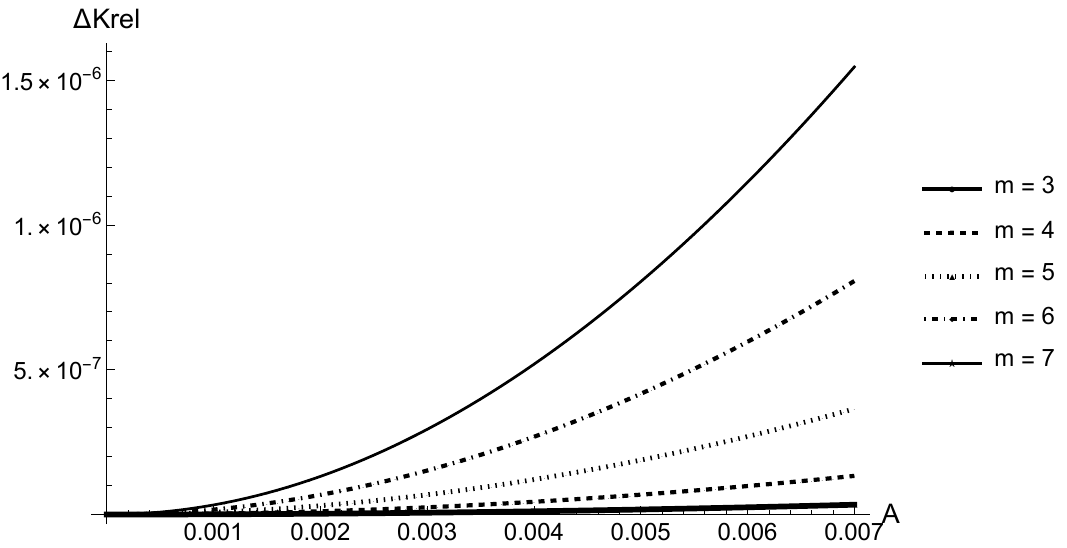}
\caption{{ Relative kinetic energy variation for several modes $Re(\zeta^{m})$. The initial transverse derivatives with respect to $u$ are $\dot{x}_{A0}=\dot{x}_{B0}=-0.0005$ and $\dot{y}_{A0}=\dot{y}_{B0}=0$.}}
\label{fig9}
\end{figure}

In the low--velocity regime, where the physical coordinate velocities remain small, the numerical analysis indicates that the dependence of the relative kinetic-energy variation on the wave amplitude is accurately described by a quartic polynomial. We consider the fit
\begin{equation}\label{model}
\Delta K_{\rm rel}\sim aA^{4}+bA^{3}+cA^{2}.
\end{equation}
For the configurations displayed in Fig.~\ref{fig9}, the corresponding fits are
\begin{widetext}
\begin{align*}
m=3:\quad
\Delta K_{\rm rel}(A)
&\approx
7.01\times10^{-4}A^{2}
-1.48\times10^{-3}A^{3}
-5.53\times10^{-3}A^{4},
\\
m=4:\quad
\Delta K_{\rm rel}(A)
&\approx
2.78\times10^{-3}A^{2}
-7.78\times10^{-3}A^{3}
-8.57\times10^{-2}A^{4},
\\
m=5:\quad
\Delta K_{\rm rel}(A)
&\approx
7.67\times10^{-3}A^{2}
-2.64\times10^{-2}A^{3}
-6.57\times10^{-1}A^{4},
\\
m=6:\quad
\Delta K_{\rm rel}(A)
&\approx
1.71\times10^{-2}A^{2}
-6.86\times10^{-2}A^{3}
-3.41A^{4},
\\
m=7:\quad
\Delta K_{\rm rel}(A)
&\approx
3.32\times10^{-2}A^{2}
-1.45\times10^{-1}A^{3}
-1.38\times10^{1}A^{4}.
\end{align*}
\end{widetext}
For all the above cases, the coefficient of determination satisfies $R^{2}>0.99$, showing that the quartic model provides an excellent description of the numerical data within the amplitude range considered.

The relevant information that can be extracted from Fig.~\ref{fig9} is not simply that the relative kinetic-energy variation becomes larger for some values of $m$, since its absolute magnitude depends on the initial placement and motion of the particles. Rather, the important point is that the coefficient of the quartic term in Eq.~\eqref{model} is strongly sensitive to the mode $m$. For the fixed initial configuration considered here, its magnitude increases with $m$, with
\begin{equation}
|a_{3}|\sim\mathcal{O}(10^{-3}),\,\,\,
|a_{4}|\sim\mathcal{O}(10^{-2}),\,\,\,
|a_{5}|\sim\mathcal{O}(10^{-1}),
\end{equation}
followed by
\begin{equation}
|a_{6}|\sim\mathcal{O}(10^{0}),\qquad
|a_{7}|\sim\mathcal{O}(10^{1}).
\end{equation}
Thus, for this particular configuration, successive polynomial modes produce an increase in the characteristic scale of the quartic coefficient. This behavior suggests that the higher-order contribution to the relative kinetic-energy memory retains quantitative information about the transverse structure of the pp--wave profile.
}

\subsection{Origin of the quartic scaling}

{ The polynomial dependence of the relative kinetic-energy variation on the amplitude of the gravitational wave may be understood analytically from the structure of the geodesic equations and from the relation between the affine derivatives and the physical coordinate velocities.}
Since the wave profile is proportional to the amplitude,
\begin{equation}
H = \mathcal{O}(A),
\end{equation}
the geodesic equations imply
{
\begin{equation}
\ddot{x}=\frac{1}{2}\partial_{x}H,
\qquad
\ddot{y}=\frac{1}{2}\partial_{y}H,
\end{equation}
and therefore the wave-induced changes in the transverse derivatives satisfy, at leading order,
\begin{equation}\label{velocitiesorder}
\Delta\dot{x},\Delta\dot{y}\sim \mathcal{O}(A).
\end{equation}
}

{ For the parametrization \eqref{kapparest} adopted in the numerical analysis, the physical coordinate velocities are, in the asymptotic region $H\rightarrow0$,
\begin{align}
V^{x}\equiv\frac{dx}{dt}
&=
\frac{2\sqrt{2}\,\dot{x}}{S+4},
\\
V^{y}\equiv\frac{dy}{dt}
&=
\frac{2\sqrt{2}\,\dot{y}}{S+4},
\\
V^{z}\equiv\frac{dz}{dt}
&=
\frac{S}{S+4},
\end{align}
where
\begin{equation}
S=\dot{x}^{2}+\dot{y}^{2}.
\end{equation}
For $S\ll1$, these expressions reduce to
\begin{align}
V^{x}&\simeq\frac{\dot{x}}{\sqrt{2}}
\left(1-\frac{S}{4}+\cdots\right),
\\
V^{y}&\simeq\frac{\dot{y}}{\sqrt{2}}
\left(1-\frac{S}{4}+\cdots\right),
\\
V^{z}&\simeq\frac{S}{4}+\mathcal{O}(S^{2}).
\label{Vzscaling}
\end{align}}

{ Correspondingly, the asymptotic kinetic energy of a single particle, Eq.~\eqref{kineticapprox}, can be expanded as
\begin{equation}
K
=
\frac{S(S+8)}{2(S+4)^{2}}
=
\frac{S}{4}
-\frac{3S^{2}}{32}
+\mathcal{O}(S^{3}).
\label{Kexpansion}
\end{equation}
An analogous order counting applies to the relative kinetic energy constructed from the differences of the physical coordinate velocities.}

{ At leading order the wave-induced transverse motion is linear in $A$. However, the complete geodesic trajectories depend on the amplitude through the positions entering $H(u,x,y)$. Consequently, the transverse response may be written as
\begin{equation}
\Delta V_{\rm rel}^{\perp}
=
A\,\mathbf{W}_{1}
+
A^{2}\,\mathbf{W}_{2}
+
A^{3}\,\mathbf{W}_{3}
+\cdots ,
\end{equation}
where the vectors $\mathbf{W}_{n}$ depend on the initial configuration, the pulse profile, and the polynomial mode. The corresponding transverse relative kinetic energy therefore contains
\begin{align}
K_{\rm rel}^{\perp}
\sim\,
&
\frac{A^{2}}{2}\,
\mathbf{W}_{1}^{2}
+
A^{3}\,
\mathbf{W}_{1}\cdot\mathbf{W}_{2}
\nonumber\\
&
+
\frac{A^{4}}{2}
\left(
\mathbf{W}_{2}^{2}
+
2\mathbf{W}_{1}\cdot\mathbf{W}_{3}
\right)
+\cdots .
\label{Ktransscaling}
\end{align}
Thus, the quadratic, cubic, and quartic terms observed numerically arise naturally when the higher-order dependence of the particle trajectories on the wave amplitude is retained. Consequently, the variation of the relative kinetic energy may be represented, to this order, by the schematic form \eqref{model}, with coefficients that depend on the initial particle configuration and on the pulse profile.}

This scaling behavior can also be understood more directly from the curvature responsible for the tidal interaction. 
{ Writing the Gaussian profile as $f(u)=A\,\bar f(u)$, the curvature components at fixed transverse position are linear in the amplitude,}
\begin{equation}
R^{i}{}_{uju}=-\frac{1}{2}\partial_i\partial_j H
           =A\,\mathcal{R}^{i}{}_{j}(u,x,y).
\end{equation}

{ Within the short-pulse approximation employed in Sec.~III B, in which the variation of the separation during the interaction is neglected at leading order, the memory tensor therefore scales as}
\begin{equation}
\mathcal{M}^{i}{}_{j}\propto A.
\end{equation}
{ Hence the leading wave-induced transverse relative velocity is of order $\mathcal{O}(A)$, consistently with Eq.~\eqref{velocitiesorder}, and its leading contribution to the relative kinetic energy is of order}
\begin{equation}
K_{\perp}\sim \mathcal{O}(A^{2}).
\end{equation}

{ An additional quartic contribution has a particularly simple origin in the longitudinal motion. From Eq.~\eqref{Vzscaling}, for particles initially at rest, or whenever the wave-induced transverse motion dominates over a sufficiently small initial transverse velocity, one has}
\begin{equation}
V^{z}\sim
\dot{x}^{2}+\dot{y}^{2},
\end{equation}
which implies
\begin{equation}
V^{z}\sim \mathcal{O}(A^{2}).
\end{equation}

{ Therefore, the longitudinal contribution to the relative kinetic energy contains a term scaling as}
\begin{equation}
K_{\parallel}\sim (\Delta V^{z})^{2}
\sim \mathcal{O}(A^{4}).
\end{equation}

{ Thus, the longitudinal motion provides a direct mechanism for generating a quartic contribution to the energy memory. The tidal interaction first produces transverse motion at leading order in the wave amplitude, while the normalization condition relates the longitudinal physical velocity nonlinearly to the transverse derivatives. At the same time, Eq.~\eqref{Ktransscaling} shows that higher-order corrections to the transverse motion may also contribute to the quartic coefficient. The observed quartic term should therefore be understood as a consequence of the combined transverse and longitudinal dynamics of the particle in the exact pp--wave geometry, rather than as originating exclusively from one component of the motion.}

The numerical results also indicate that the magnitude of the quartic coefficient in Eq.~\eqref{model} depends on the { polynomial mode $m$}. This behavior can be understood from the spatial structure of the pp--wave profile.

Since the transverse motion is driven by the spatial derivatives of $H$, { the leading wave-induced transverse derivatives} scale as
\begin{equation}
\Delta\dot{x},\Delta\dot{y}\sim A\,\nabla\Phi_m .
\end{equation}

For the modes in polar coordinates $(r,\varphi)$ we have $\Phi_m=r^m\cos(m\varphi)$, thus one finds
\begin{equation}
|\nabla\Phi_m|^2=m^2 r^{2m-2}.
\end{equation}
Hence { the leading quadratic combination of the wave-induced transverse derivatives} scales as
\begin{equation}
(\Delta\dot{x})^2+(\Delta\dot{y})^2
\sim A^2 m^2 r^{2m-2}.
\end{equation}

{ The longitudinal contribution identified above is quadratic in this quantity. Consequently, its contribution to the coefficient of the $A^4$ term inherits the approximate spatial scaling}
\begin{equation}
a_m^{\parallel}
\propto
|\nabla\Phi_m|^4
\sim
m^4 r^{4m-4}.
\end{equation}
{ This relation is a leading-order scaling estimate, since the full quartic coefficient also contains higher-order contributions from the transverse trajectories and depends on the initial particle configuration.}

{ For the configuration considered in Fig.~\ref{fig9}, in which the particles are initially located close to $r=1$, this scaling provides a qualitative explanation for the increase of the magnitude of the quartic coefficient with $m$. In particular, near $r=1$ the leading spatial factor behaves approximately as $m^{4}$.}

{ Ergo, the growth observed numerically is related to the increasingly strong transverse gradients of the polynomial profiles $\Phi_m$. Since these gradients control the tidal acceleration and the resulting particle motion, the coefficient of the $A^4$ term retains quantitative information about the transverse structure of the pp--wave mode.}

\section{Conclusions}\label{conc}

In this article we investigated the memory effect for test particles interacting with a pp--wave Gaussian pulse. We { extended the analysis to higher-order polynomial pp--wave modes beyond the usual quadratic profiles}, finding a persistent memory effect for both the coordinates and the velocities. In this way we extend the literature on the subject developed by Zhang and collaborators \cite{zhang2017memory,zhang2018ion,zhang2018memory,zhang2018velocity,zhang2020scaling}. 

Given the persistence of the velocity memory effect, we also extended the results of \cite{maluf2018kinetic,maluf2018plane,maluf2018variations} to higher modes, showing that the energy memory effect persists for all the modes considered here. { In particular, the higher-order dependence of the relative kinetic-energy variation on the wave amplitude reflects the combined transverse and longitudinal particle dynamics in the exact pp--wave geometry.}

In order to mitigate the possibility that the observed memory effect could be merely a coordinate artifact, we reexamined the problem by considering the relative motion between two test particles, whether neighboring or not. Within this framework we verified that the effect persists, reinforcing the physical character of the phenomenon.

{ Since the relative kinetic energy of the particles may increase or decrease depending on the wave amplitude, we investigated its dependence on the amplitude.} Our results indicate that smaller amplitudes tend to decrease { the relative kinetic energy} whereas larger amplitudes tend to increase it. In the regime of small velocities we found a quartic polynomial dependence on the amplitude, which was subsequently explained analytically.

{ Within the approximations of small velocities and short pulses, we also showed that the relative kinetic-energy memory depends on the history of the curvature tensor, indicating that the effect is intrinsically global rather than local. The leading transverse response produces a quadratic contribution in the wave amplitude, while the longitudinal motion provides a direct mechanism for generating a quartic contribution. Higher-order corrections to the transverse dynamics may also contribute to the quartic term. Furthermore, we showed that, for a fixed value of $\kappa$, the longitudinal motion is determined by the transverse one, so that the particle trajectories can be characterized by the four transverse initial parameters $x_{0},y_{0},\dot{x}_{0},\dot{y}_{0}$.}

{ The results obtained here should be interpreted primarily as a theoretical characterization of the memory effect in exact pp--wave geometries. In particular, the dependence of the relative kinetic-energy variation on the parameter $m$ provides information about how the different transverse structures of polynomial pp--wave profiles affect the motion of massive test particles. Establishing a connection between these results and astrophysical gravitational-wave signals would require a separate analysis relating such pp--wave profiles to realistic source-generated waveforms, which lies beyond the scope of the present work. The absence of such a connection in the present analysis should not be interpreted as excluding a possible phenomenological relevance of these profiles, but only as indicating that no such source correspondence is established here.}

We note an interesting qualitative connection with recent results on approximate models for gravitational memory \cite{zhao2026approximate}. In those models, it was shown that discrete critical amplitudes select solutions with vanishing velocity memory through global matching conditions, reflecting a cancellation of the tidal interaction along the particle's history. In the present work, although the relevant observable is the asymptotic variation of the kinetic energy rather than the final velocity itself, we observe that specific amplitudes may lead to a vanishing net energy transfer. This suggests that both phenomena may be manifestations of a common underlying mechanism, namely a global cancellation encoded in the associated Sturm--Liouville dynamics, albeit at the level of different observables. In the low-velocity regime, where the kinetic energy is directly related to the particle's velocity, this correspondence becomes particularly suggestive.

The qualitative resemblance with the Landau damping effect is also noteworthy. When two thermodynamical systems $A$ and $B$ are placed in thermal contact, the direction of the energy flow (heat) depends not only on the individual temperatures of the systems but on their relative values. Analogously, if one treats the particle and the pp--wave as two interacting systems, the energy transfer between them may depend on their relative properties rather than on their individual energies alone, a qualitative behavior that was observed in the present analysis.

If one considers a system composed of many test particles, it is possible to define statistically an effective temperature associated with their velocity distribution. On the other hand, if a notion of gravitational energy is attributed to the pp--wave, as discussed in Refs.~\cite{maluf2008energy,formiga2023gravitational,formiga2025angular,costa2025gravitational}, one may speculate about the possibility of associating an effective thermodynamical temperature to the wave itself, related to its energy content and therefore to its amplitude. In this interpretation the interaction between the wave and the particles could be viewed, at least qualitatively, as a thermodynamical exchange process. In particular, for combinations of initial conditions and amplitudes that produce no net energy transfer, as illustrated in Fig.~\ref{fig8}, one may regard the composite system (gravitational wave plus particles) as being in a state analogous to thermal equilibrium. Future investigations along these lines may provide further insight into possible thermodynamical aspects of the gravitational field as probed by the motion of real particles.


\section*{Acknowledgments}

The development of this research was supported by the Research and Graduate Support Program (PROPESQ/UFNT), under Grant No.~038/2025. F. L. Carneiro acknowledges financial support from the Brazilian National Council for Scientific and Technological Development (CNPq), Brazil, through a Productivity Fellowship (PQ), Grant No. 303574/2026-7.


\section*{Data Availability}

There are no publicly available research data or software supporting this manuscript. Requests for further information or data should be sent to the authors.

\end{document}